\documentclass[12pt,a4paper]{article}

\usepackage[utf8]{inputenc}
\usepackage[nottoc]{tocbibind}

\usepackage{amsmath,amssymb}				
\usepackage{amsfonts}				
\RequirePackage{bbm}
\usepackage[thref,amsmath,thmmarks]{ntheorem}	
\usepackage{prettyref}				
\usepackage{multicol}
\usepackage{paralist,epsfig}				
\usepackage{tikz}
\usetikzlibrary{calc,intersections,arrows}

\usepackage{environ}
\makeatletter
\newsavebox{\measure@tikzpicture}
\NewEnviron{scaletikzpicturetowidth}[1]{%
  \def\tikz@width{#1}%
  \begin{lrbox}{\measure@tikzpicture}%
  \BODY
  \end{lrbox}%
  \pgfmathparse{#1/\wd\measure@tikzpicture}%
  \BODY
}
\makeatother

\usepackage{url}
\usepackage[colorlinks=true]{hyperref}
\usepackage{tensor}
\usepackage{lmodern}
\usepackage{textcomp}
\usepackage{graphicx}

\newcommand {\bC}{{\mathbb C}}

\newcommand {\bN}{{\mathbb N}}

\newcommand {\bR}{{\mathbb R}}

\newcommand {\cC}{{\mathcal C}}
\newcommand {\cD}{{\mathcal D}}

\newcommand {\cF}{{\mathcal F}}

\newcommand {\cH}{{\mathcal H}}

\newcommand {\cJ}{{\mathcal J}}
\newcommand {\cK}{{\mathcal K}}

\newcommand {\cM}{{\mathcal M}}

\newcommand {\cO}{{\mathcal O}}
\newcommand {\cP}{{\mathcal P}}

\newcommand {\bfC}{C}

\newcommand {\bsd}{\boldsymbol{d}}
\newcommand {\bsi}{\boldsymbol{i}}

\newcommand {\dist}{\operatorname{dist}}

\newcommand {\Sing}{\operatorname{Sing}}
\newcommand {\Coll}{\operatorname{Coll}}

\newcommand {\Trans}{\operatorname{Trans}}
\newcommand {\Wand}{\operatorname{Wand}}

\newcommand {\hM}{{\widehat{M}}}
\newcommand {\hP}{{\widehat{P}}}

\theoremnumbering{arabic}

\theoremstyle{plain}
\RequirePackage{latexsym}
\theorembodyfont{\itshape}
\theoremheaderfont{\normalfont\bfseries}
\theoremseparator{}

\newtheorem{Theorem}{Theorem}[section]
\newtheorem{Proposition}[Theorem]{Proposition}
\newtheorem{Lemma}[Theorem]{Lemma}
\newtheorem{Corollary}[Theorem]{Corollary}

\newtheorem{Example}[Theorem]{Example}
\newtheorem{Remark}[Theorem]{Remark}

\newtheorem{Definition}[Theorem]{Definition}

\theoremstyle{nonumberplain}
\theorembodyfont{\upshape}

\theoremheaderfont{\itshape}
\theorembodyfont{\normalfont}
\theoremsymbol{\ensuremath{_\Box}}
\RequirePackage{amssymb}
\qedsymbol{\ensuremath{_\Box}}

\theoremclass{LaTeX}

\numberwithin{equation}{section}
\newcommand {\eh}{{\textstyle \frac{1}{2}}}
\newcommand{\qmbox}[1]{\quad\mbox{#1}\quad}
\newcommand{\idty}{{\rm 1\mskip-4mu l}} 

\begin{document}

\title{Improbability of Collisions in $n$-Body Systems}
\author{Stefan Fleischer and Andreas Knauf\thanks{
Department of Mathematics,
Friedrich-Alexander-University Erlangen-N\"urnberg,
Cauerstr.\ 11, D-91058 Erlangen, 
Germany, \texttt{fleischer,knauf@math.fau.de}} 
}
\date{\today}
\maketitle
\begin{abstract}
For a wide class of two-body interactions, 
including standard examples like gravitational or Coulomb fields,
we show that collision orbits in $n$-body systems are of Liouville measure zero for all energies. We use techniques from symplectic geometry to relate the volume of the union of collision orbits to the area of  
Poincar\'{e} surfaces surrounding the collision set.
\end{abstract}
%
\section{Introduction}
%
Consider as a primary example the motion of $n\in\bN$ particles with masses 
$m_1,\ldots,m_n>0$ due to Newton's law of gravitation:
\begin{equation}
\label{eq:Newton}
m_i \ddot q_i = \sum_{j\in\{1,\ldots,n\}\setminus\{i\}} \frac{m_j m_i (q_j - q_i)}{\| q_j - q_i\|^3} 
\qquad (i=1,\ldots,n).
\end{equation}
Here we have set the scale of time in a way that the gravitational constant becomes $1$. For $n>1$
the flow of this ordinary differential equation is obviously not complete; for instance consider two 
particles, whose initial velocity vectors are pointing exactly towards each other -- they will collide in finite time.
Phase space points respectively their positive semi-orbits are called {\em singular}, if their maximal time interval
of existence (for non-negative times) is finite. A singularity is called a {\em collision}, 
if all particles have limit positions 
in configuration space, as time approaches singular time, see Section \ref{sec:Statement} for further details.

In the papers \cite{Saari_1971,Saari_1973}, {\sc Saari} has shown the improbability of collisions, meaning 
that all collision points define a subset of phase space of Lebesgue measure zero. 
His techniques can be used to generalize this result to a class of homogeneous force fields including 
the gravitational case. As he points out, however, his bounds are not optimal.

Here, we generalize the result to an even wider class of potentials, also implementing optimal bounds. 
By giving up assumptions like homogeneity of the force field, we cannot rely on certain 
arguments any longer, for example arising from the Lagrange-Jacobi-Identity.
Instead, we employ geometric techniques:

The first is based on a decomposition of configuration space,
invented by Gian-Michele Graf in showing asymptotic completeness of quantum scattering.

The second comes from symplectic geometry: after defining an appropriate sequence of 
hypersurfaces surrounding the collision set, we can relate their surface area to the volume of the set of 
initial points, whose orbits are passing through the surfaces.
The technical aspects of this method have been worked out in \cite{Fleischer_Knauf_2018}.

The outline of the paper is as follows: in Section \ref{sec:Statement} we define our class of
admissible potentials and state the main result, Theorem \ref{thm:ImprobColl}. 
Section \ref{sec:Definition:C} presents an adapted partition of configuration space
and a sequence of hypersurfaces onto which the Poincar\'{e} surfaces project.
In Section \ref{sec:Definition:P}, by building on the main technique introduced in \cite{Fleischer_Knauf_2018}, 
we give the definition of a sequence of Poincar\'{e} surfaces.
Then we estimate their symplectic (as opposed to Riemannian) volume in Proposition \ref{prop:area}.
In Proposition \ref{prop:int:kin} of Section \ref{sec:Time:Int} we prove finiteness of the time integral of
kinetic energy.
In Section \ref{sec:Hitting:P} the Poincar\'{e} surfaces are then shown to eventually be hit by every 
collision orbit (Proposition \ref{prop:hit}).

In combination, Propositions \ref{prop:area} and \ref{prop:hit}, together with Theorem A
of \cite{Fleischer_Knauf_2018}, imply our main result, Theorem \ref{thm:ImprobColl}.

In the final section \ref{sec:Addenda} we indicate the minimal changes that need to be done in order to 
prove analogous theorems 
\begin{enumerate}
\item [$\bullet$]
for multiple collisions on the line ($d=1$), or
\item [$\bullet$]
in the presence of fixed centers.
\end{enumerate}

\medskip
{\bf General Notation:} 
We point out that we use $C$ as a positive constant, that only depends on fixed system parameters like 
number of particles or their masses. Its value may change with every usage, even within the same set 
of equations.

\section{Statement of the Main Result}
\label{sec:Statement}

We consider the motion of $n\geq 2$ particles with respective masses $m_1,\ldots,m_n$ in $d\geq 2$ dimensions. Thus the (a priori) configuration space is given by
\begin{equation*}
 M := M_1 \oplus \ldots \oplus M_n,
\end{equation*}
with $M_k := \bR^d$ for $k\in N := \{1,\ldots,n\}$. We write the elements of $M$ in the form
\begin{equation*}
 q = \left(q_1,\ldots,q_n \right). 
\end{equation*}
Using the canonical identification $T^* M = M \times M^*$ with $M^*$ being the vector space dual to $M$, 
we write the elements of the cotangent bundle $T^* M$ in the form
\begin{equation*}
 x=(q,p)=(q_1,\ldots,q_n,p_1,\ldots,p_n).
\end{equation*}
The inner product on $M$ defined by
\begin{equation}
\label{inner:prod:M}
 \langle \cdot,\cdot \rangle_M : M \times M \rightarrow \bR \qmbox{,} 
 \langle q,q'\rangle_M := \langle q, \cM q' \rangle, 
\end{equation}
where $\langle \cdot,\cdot \rangle$ is the Euclidean inner product on $M$ and
\begin{equation*}
 \cM := \operatorname{diag}(m_1,\ldots,m_n) \otimes \mathbbm{1}_d
\end{equation*}
is the scaling according to the masses, induces an inner product on $M^*$ via
\begin{equation*}
 \langle \cdot,\cdot \rangle_{M^*} : M^* \times M^* \rightarrow \bR \qmbox{,}  
 \langle p,p'\rangle_{M^*} := \langle p, \cM^{-1} p' \rangle
\end{equation*}
as well as an inner product on the cotangent bundle $T^* M$ via
\begin{equation}
\label{TsM:product}
 \langle \cdot,\cdot \rangle_{T^* M} : T^*M \times T^*M \rightarrow \bR, \quad \langle x,x'\rangle_{T^*M} := \langle q, q' \rangle_M + \langle p, p' \rangle_{M^*}.
\end{equation}
By this, we get a Riemannian manifold $(T^*M, g)$ with $g\equiv \langle \cdot,\cdot\rangle_{T^*M}$, the Riemannian volume form of which is given by the symplectic volume form $\Omega_{nd}$, with
\begin{equation}
\label{eq:DefOmega}
 \Omega_k:= \frac{(-1)^{\lfloor k/2 \rfloor}}{k!}\omega^{\wedge k}\qquad(k=1,\ldots,nd).
\end{equation}
Here $\omega$
is the canonical symplectic form on $T^*M$.
$\cM$ induces an identification $T^* M \cong \bR^{2nd}$, the Riemannian volume form is equal to Lebesgue measure $\lambda^{2nd}$. We will use that $(T^* M,g,\omega)$ is a K\"ahler manifold.

The force field defining the motion consists of two-body interactions. 
We write
\begin{equation}
\label{eq:DefDelta}
\Delta := \{ q\in M \ | \  \mbox{there exist $i\neq j\in N$ with } q_i = q_j \}
\end{equation}
for the {\em collision set}. Thus the (actual) configuration space $\hM$ is defined by
\begin{equation*}
 \hM := M \setminus \Delta.
\end{equation*}
On phase space $\hP := T^* \hM$, the Hamiltonian function is defined by 
\begin{equation}
 \label{eq:DefH}
H \in C^2(\hP,\bR), \qquad H(q,p) := K(p) + V(q).
\end{equation}
Here,
\begin{equation}
 \label{eq:DefK}
 K:T^*M \rightarrow \bR, \quad K(q,p) \equiv K(p) = \textstyle{\frac12} \langle p,p \rangle_{M^*} 
 = \sum_{k \in N} \frac{\|p_k\|^2}{2 m_k}
\end{equation}
is the kinetic energy and $V:\hM \rightarrow \bR$ is the potential; we assume the potential to be of the form
\begin{equation}
\label{eq:DefV}
 V(q) = \sum_{i < j \in N} V_{i,j}(q_i-q_j) 
\end{equation}
with two-body potentials $V_{i,j}\in C^2(\bR^d\setminus\{0\},\bR)$. 
For simplifying notation, we write
\begin{equation}
 V_{j,i}(q) = V_{i,j} (-q) \qquad (1\leq i < j \leq n, \ q \in \bR^d\setminus\{ 0\}).
\end{equation}
The Hamiltonian vector field $X_H $ is defined by the equation 
$\bsi_{X_H} \omega = \bsd H$, where $\bsi$ is the inner product and $\bsd$ is the exterior derivative.
So it is continuously differentiable, and the Hamiltonian differential equation $\dot x = X_H(x)$ has local
solutions. In coordinates, the differential equation is given by
\begin{equation*}
\dot{q}_k = \frac{p_k}{m_k}\qmbox{,} \dot{p}_k = -\sum_{i \in N\setminus\{k\}} \nabla V_{k,i} (q_k - q_i) 
\qquad (k \in N).
\end{equation*}

\begin{Definition}\label{def:admissible}
We call the potential $V$ {\bf admissible}, if 
$\,\lim_{\|q\|\to\infty}V_{i,j}(q)=0$, there exists an $\alpha\in(0,2)$ such that 
\begin{equation}
\label{D2V}
D^2V_{i,j}(q)=\cO(\|q\|^{-\alpha-2})\qquad(\|q\|\le 1),
\end{equation}
and for some $C_{V}>0$ either
\begin{enumerate}
\item 
for suitable $Z_{i,j}\in \bR$ 
\begin{equation}
\label{eq:admiss1}
\Big|\Big\langle \frac{q}{\|q\|},\nabla V_{i,j}(q)\Big\rangle +\alpha\frac{Z_{i,j}} {\|q\|^{\alpha+1}}\Big| \le C_{V} 
\ \qquad (\|q\| \in(0, 1],\ 1\leq i < j \leq n)
\end{equation}
\item 
or the $V_{i,j}$ are bounded above, and, with $W_-(q):=\max(-W(q),0)$,
\begin{equation}
\label{eq:admiss2}
\langle q,\nabla V_{i,j}(q)\rangle \le C_{V} +\alpha\,(V_{i,j})_-(q) \ \qquad (\|q\| \in(0, 1],\ 1\leq i < j \leq n).
\end{equation}
\end{enumerate}
\end{Definition}

\begin{Example}[Admissible Potentials]\quad\\[-6mm]
\begin{enumerate}[1.]
\item 
An important class of admissible potentials consists of the homogeneous potentials
\begin{equation*}
 V_{i,j}(q) = \frac{Z_{i,j}}{\| q\|^{\alpha_{i,j}}},
\end{equation*}
with $Z_{i,j} \in \bR$ and  $\alpha_{i,j} =\alpha \in (0,2)$, or with $Z_{i,j} <0$ and $\alpha_{i,j} \in (0,2)$. 

In particular, this includes the cases of gravitational and Coulomb force fields.\footnote{
With $\alpha_{i,j} =1$ and $Z_{i,j} := -m_i m_j$ for all $1 \leq i < j \leq n$, we get the case of interaction 
due to gravitation, cf. \eqref{eq:Newton}. 
With $Z_{i,j} := \rho_i \rho_j$ (here, the particles' charges $\rho_k\in\bR$ take over the role of the particles' masses) we get the case of interaction due to static electrical charge.}

Note that collisions of three or more charged particles are possible, although some of them then necessarily 
repel each other. 

Perhaps the easiest case is the one of a particle with charge $\rho_1>0$ resting at the
origin and two particles with masses $m_2=m_3$ and charges $\rho_2=\rho_3\in (-4\rho_1,0)$,
with positions $q_2=-q_3$ and momenta $p_2=-p_3$.
\item 
\eqref{eq:admiss1} also includes the physically important case of Yukawa potentials
\[V_{i,j}(q)=Z_{i,j}\frac{\exp(-m_{i,j}\|q\|)}{\|q\|}\qquad(Z_{i,j}\in\bR,\, m_{i,j}>0).\]
Then
\[\Big|\Big\langle \frac{q}{\|q\|},\nabla V_{i,j}(q)\Big\rangle +\alpha\frac{Z_{i,j}} {\|q\|^{\alpha+1}}\Big|
=\frac{Z_{i,j}}{\|q\|^2}(1-e^{-m_{i,j}\|q\|})(m_{i,j}\|q\|+1)=\cO(1).\mbox{$\Diamond$} \]
\end{enumerate}
\end{Example}
\begin{Remark}[Admissible Potentials] \label{rem:condition1}\quad\\
Aside for allowing for positive and negative singularities, Condition \eqref{eq:admiss1}  is much stricter than
\eqref{eq:admiss2}, as up to a constant the potentials $V_{i,j}$ are homogeneous near zero:
\eqref{eq:admiss1} entails for $q\in\bR^d\!\setminus\!\{0\}$ with $\|q\|\le 1$  
\begin{align*}
V_{i,j}(q)&=V_{i,j}(q/\|q\|)-\int_1^{\|q\|^{-1}}\frac{d}{ds} V_{i,j}(sq)\,ds\\
&\leq V_{i,j}(q/\|q\|)+\int_1^{\|q\|^{-1}} \left( \frac{Z_{i,j}}{\|q\|^\alpha}\alpha s^{-(\alpha+1)}+C_V\|q\|\right) \,ds\\
&= V_{i,j}(q/\|q\|) +C_V(1-\|q\|) -Z_{i,j}+  \frac{Z_{i,j}}{\|q\|^\alpha},
\end{align*}
and similarly for the other direction of the inequality.
So by compactness of $S^{d-1}$
\[\left|V_{i,j}(q)-\frac{Z_{i,j}}{\|q\|^\alpha}\right| \le |V_{i,j}(q/\|q\|)| + |Z_{i,j}|+C_V(1-\|q\|) = \cO(1).
\qquad \qquad \quad \mbox{$\Diamond$}\]
For homogeneous attracting potentials proofs simplify, since one can make use of 
the results of {\sc Pollard} and {\sc Saari} in \cite{Pollard_Saari_1968}.
\end{Remark}

Going back to the Hamiltonian system as defined above, the corresponding Hamiltonian flow $\Phi:D \rightarrow \hP$ uniquely exists on a maximal neighborhood $D \subseteq \bR \times \hP$ of $\{0\}\times \hP$ in extended phase space; we have $\Phi\in C^1(D,\hP)$. Shortly, we write
\begin{equation*}
 \Phi(t,x) = \Phi_t(x) = (q(t,x),p(t,x)) = (q(t),p(t)) \qquad((t,x)\in D),
\end{equation*}
the latter if there is no ambiguity concerning the initial condition $x\in \hP$.

The flow's domain of definition is of the form
\begin{equation}
\label{eq:DefD}
 D = \left\{ (t,x) \in \bR \times \hP \ \left| \  T^-(x) < t < T^+(x) \right.\right\}
\end{equation}
with the escape time $T = T^+: \hP \rightarrow (0,\infty]$; by reversibility of $X_H$ we have $T^-(q,p) = -T^+(q,-p)$. Additionally, $T^+$ is lower semi-continuous.

By
\begin{equation}
\label{eq:Sing}
 \Sing := \left\{ x \in \hP \ | \  T(x) < \infty \right\},
\end{equation}
we denote the set of phase space points experiencing a singularity.

In celestial mechanics, it is a well known fact due to Painlev\'{e}, that a singularity occurs if and only if the minimal particle distance converges to zero. 
As a first result, we point out that this still holds in our more general setting of two-body interactions, since the classical proof can be applied.
For this purpose, let
\begin{equation}
\label{q:min}
q_{\min}:\hP\to (0,\infty)\qmbox{,} (q,p)\mapsto \min\{\| q_i - q_j \| \mid i\neq j \in N\}.
\end{equation}
be the minimal distance of particles.
Then we get:
\begin{Theorem}[Painlev\'{e}]
\label{thm:Painleve}
Let $x \in \Sing$. Then
$\lim_{t \nearrow T(x)} q_{\min}\circ \Phi_t(x) = 0$.
\end{Theorem}
{\bf Proof: }
Otherwise, there exist $\delta>0$ and a sequence of monotonically increasing times $(s_j)_{j\in\bN}$ with 
$\lim_{j\rightarrow\infty}s_j = T(x)$ and $\dist(q(s_j),\Delta) > \delta$. 
By assumption,  the potential is bounded below on the domain
\begin{equation*}
 U := \left\{ q \in \hM \ \left| \ \dist(q,\Delta) \geq \tfrac{\delta}2 \right.\right\},
\end{equation*}
that is $V_{\min} := \inf_{q\in U} V(q) \in \bR$. 
Thus by conservation of energy $E := H(x)$, as long as $q(t) \in U$, 
velocity is bounded above, namely 
\[\|\dot{q}(t)\| \leq v_{\max}:=\sqrt{2m_{\max}(E-V_{\min})}\] 
with $m_{\max}:=\max\{m_1,\ldots,m_n\}$. 
So for all $j\in \bN$ the solution can be extended at least up to $s_j + \tfrac{\delta}{2v_{\max}}$.
This contradicts the assumption $\lim_{j\rightarrow\infty}s_j = T(x)$.
\hfill$\Box$\\[2mm]
Within this work, we are particularly interested in those singularities, which have limit positions in configuration space at singular time, and call them collision singularities:
\begin{equation}
\label{eq:Coll}
 \Coll := \left\{ x \in \Sing \ \left| \  \lim_{t \nearrow T(x)} q(t,x) \mbox{ exists (in $M$)} \right.\right\}.
\end{equation}

Furthermore, we restrict considerations to the energy surfaces   
\[\Sigma_E := H^{-1}(E)\qquad (E\in\bR).\] 
Since $\Sing$ is a subset of the open domain consisting of all non-equilibrium points in $\hP$, 
we can without loss of generality assume that every $E\in\bR$ is a regular value of $H$. 
Thus, $\imath_E:\Sigma_E\to \hP$ is a codimension one submanifold (if non-empty). 
We write $\Coll_E := \Coll \cap \Sigma_E$. 

There is a $(2nd - 1)$--form $\sigma$ on 
phase space $\hP$ with $dH \wedge \sigma = \Omega_{nd}$, see Remark~1.4 of \cite{Fleischer_Knauf_2018}.
Although $\sigma$ is not fixed by that property, its pull-back $\sigma_E:=\imath_E^*\sigma$
is a uniquely defined volume form on that energy surface, invariant under the restricted flow. 
We denote by $\sigma_E$, too the corresponding {\em Liouville measure} on~$\Sigma_E$.

Now we can state our main result:

\begin{Theorem}
\label{thm:ImprobColl}
For all $n\in\bN$, $d\ge2$ and $E\in\bR$ 
the set $\Coll_E$ of phase space points leading to a collision has Liouville measure zero, provided $V$ is admissible.
\end{Theorem}
By integration with respect to total energy $E$ it follows that the Lebesgue measure $\lambda^{2dn}(\Coll)$
of the collision set in phase space $\hP$ vanishes, too.

\section{Partitioning Configuration Space}
\label{sec:Definition:C}

\paragraph{Cluster Coordinates}\quad\\
%
We now introduce coordinates, that (notationally) link certain subgroups of particles together, in the form of so-called \emph{clusters}.
The {\em external} cluster coordinates then describe the motion of the cluster as a whole, 
whereas the {\em internal} ones describe each particle's motion within its cluster.\\ 
We begin with some standard notions: 
\begin{Definition}\quad\\[-6mm]
\begin{enumerate}[$\bullet$]
\item
A \textbf{set partition} or \textbf{cluster decomposition} of $N$ is a set $\cC:=\{C_1,\ldots,C_k\}$ of
\textbf{blocks} or \textbf{clusters} $\emptyset\neq C_\ell\subseteq N$ such that\vspace*{-2mm}
\[
\bigcup_{\ell=1}^kC_\ell=N \qmbox{and}
C_\ell\cap C_m=\emptyset\ \mbox{ for } \ell\ne m\, .
\]
We denote by $\sim_\cC$ (or $\sim$, if there is no ambiguity) the equivalence relation on $ N$ induced by $\cC$; the corresponding equivalence classes are denoted by $[\cdot ]_\cC$.
\item
The \textbf{lattice of partitions} $\cP(N)$ is the set of cluster
decompositions $\cC$ of $N$, partially ordered by
\textbf{refinement}, i.e., 
\[
\cC=\{C_1,\ldots,C_k\}\preccurlyeq \{D_1,\ldots,D_\ell\}=\cD\, ,
\] 
if  
$C_m\subseteq D_{\pi(m)}$ for an appropriate mapping 
$\pi:\{1,\ldots,k\}\to\{1,\ldots,\ell\}$. 
In this case, $\cC$ is called  \textbf{finer} than $\cD$ and $\cD$
\textbf{coarser} than $\cC$.\\
The unique finest and coarsest elements of $\cP(N)$ are
\[\cC_{\min} := \big\{ \{1\}, \ldots, \{n\} \big\} \qmbox{and} \cC_{\max} := \{N\}=\big\{\{1,\ldots,n\}\big\},\]
respectively.
By $\cP_0( N)$ we denote the set of all decompositions without the finest one, i.e. $\cP_0( N) := \cP( N) \setminus \{ \cC_{\min} \}$.
\item
The  \textbf{rank} of $\cC\in\cP(N)$ is the number  $|\cC|$ of its
blocks. 
\item
The  \textbf{join} of $\cC$ and $\cD\in\cP(N)$, denoted as  
$\cC\vee\cD$, is the finest cluster decomposition that is coarser than
both $\cC$ and $\cD$.
\end{enumerate}

\end{Definition}

We use partitions to decompose configuration space:
given a subset $\emptyset \neq \bfC \subseteq  N$, we define the corresponding \emph{collision set} as
\begin{equation*}
\Delta_C^E := \left\{ q \in M \ | \ q_i = q_j \mbox{ if } i,j \in C \right\},
\end{equation*}
and for a cluster decomposition $\cC$ we define the \emph{$\cC$-collision subspace}
\begin{equation}
\label{eq:DefDeltaE}
\Delta_\cC^E := \left\{ q \in M \ | \ q_i = q_j \mbox{ if } [i]_\cC = [j]_\cC \right\} = \bigcap_{C\in\cC} \Delta_C^E.
\end{equation}
By $\Pi_C^E$ we denote the $\cM$-orthogonal projection onto the subspace $\Delta_C^E$, 
and we denote the complementary projection $\idty_\cM - \Pi_C^E$ by $\Pi_C^I$.
Accordingly, we denote the projection onto $\Delta_\cC^E$ by $\Pi_\cC^E := \prod_{C\in \cC} \Pi_C^E$, 
and the complementary projection by $\Pi_\cC^I = \idty_\cM - \Pi_\cC^E = \sum_{C\in \cC} \Pi_C^I$.
The image of $\Pi_C^I$ then is given by
\begin{equation*}
\Delta_C^I := 
\left\{ q \in M \ \left| \ \sum_{i\in C} m_i q_i = 0,\ \forall \, i \in N\!\setminus\! C: \, q_i = 0 \;   \right.\right\},
\end{equation*}
the image of $\Pi_\cC^I$ is given by
\begin{equation*} 
\Delta_\cC^I := \left\{ q \in M \ \left| \ \forall\,C\in \cC: \sum_{i \in C} m_i q_i = 0  \right.\right\} = 
\bigoplus_{C\in \cC} \Delta_C^I. 
\end{equation*}
In particular, $\Delta_{\cC_{\min}}^E=M$. Regarding the dimensions of these subspaces, we have
\begin{align}
\dim( \Delta_\cC^E ) &= d \left( n - \sum_{C \in \cC} ( |C| -1) \right) = d | \cC |\,, \label{eq:DimDeltaE} \\
\dim( \Delta_\cC^I ) &= d \sum_{C \in \cC} ( |C| - 1) = d (n - | \cC | )\,. \label{eq:DimDeltaI} 
\end{align}
Thus we get a $\cM$-orthogonal decomposition
\begin{equation}
\label{M:E:I}
M = \Delta_\cC^E \oplus \bigoplus_{C\in \cC} \Delta_C^I.
\end{equation}
For a nonempty subset $C\subseteq N$ we define the 
{\em cluster mass\,}, {\em cluster barycenter}  and {\em cluster momentum} of $C$ by
\[m_C := \sum_{j\in C} m_j \qmbox{,}
  q_C := \frac{1}{m_C} \sum_{j\in C} m_j q_j \qmbox{and} p_C:=\sum_{i\in C}p_i.\]
In particular $m_N$ equals the 
{\em total mass} of the particle system.
Then for the partitions $\cC\in \cP(N)$ the $i$--th component of the 
cluster projection $q^E_\cC := \Pi^E_\cC(q)$ is given by the barycenter
\begin{equation}
\label{cl:bar}
\big(q^E_\cC\big)_i = q_{[i]_\cC} \qquad(i\in N)
\end{equation}
of its cluster. Similarly for $q^I_\cC := \Pi^I_\cC(q)$,
\begin{equation}
\label{Deltaqi}
\big(q^I_\cC\big)_i = q_i-q_{[i]_\cC} \qquad (i\in N)
\end{equation}
is its distance from the barycenter. 

Join of partitions corresponds to intersection of collision subspaces:
\[\Delta_\cC^{E}\cap \Delta_\cD^{E} = \Delta_{\cC\vee\cD}^{E}
\qquad\bigl(\cC,\cD\in \cP(N)\bigr).
\]
So the mutually disjoint sets
\begin{equation}
\label{Xi:Null}
\Xi_\cC^{(0)} := \Delta_\cC^{E}\Big\backslash \
\mbox{{\normalsize $\bigcup\limits_{\cD\succneqq\cC}$}}
\Delta_\cD^{E}\qquad\bigl(\cC\in \cP(N)\bigr),
\end{equation}
form a set partition of $M$, with $\Xi_{\cC_{\min}}^{(0)}=\widehat{M}$. 
Note that we can write $\Xi^{(0)}_\cC$ as
$\Xi^{(0)}_\cC = \{ q \in M \mid q_i = q_j \mbox{ if and only if  }i \sim_\cC j \}$.
Based on this, we partition the collision set $\Coll\subseteq \widehat{P}$ uniquely into clusters by 
\[
{\rm SP}:\Coll\to\cP_0(N) \qmbox{,} 
\lim_{t\nearrow T(x)}q(t,x)\in\Xi_{{\rm SP}(x)}^{(0)}\,.
\]

\paragraph{The Graf Partition}\quad\\
This partition, introduced by G.-M.\ Graf,  relies on the  \emph{(mean) moment of inertia}\[
J:M\to\bR\qmbox{,} J(q)=\|q\|^2_{{\!\mbox{\tiny $\cM$}}}
=\sum_{k=1}^n m_k\|q_k\|^2\,,
\]
see {\sc Derezi\'{n}ski} and {\sc G\'{e}rard} \cite[Chapter 5]{Derezinski_Gerard_1997}, 
and \cite[Chapter 12.6]{Kn18}.
\begin{Lemma}  \label{lem:CD:comparable}
In the cluster decomposition  $\cC\in \cP(N)$, $J$ is of the form 
\begin{equation}
\label{Z8}
J=J_\cC^E+J_\cC^I \qmbox{with} J_\cC^E:=J\circ\Pi_\cC^E \qmbox{and} J_\cC^I:=
J\circ\Pi_\cC^I\,.
\end{equation}
For $\cC\preccurlyeq\cD$, $J_\cC^E-J_\cD^E\ge 0$ is a quadratic form with
index of inertia $d (|\cC|-|\cD|)$.
\end{Lemma}
\textbf{Proof:}
Indeed, $J(q)=\left\langle(\Pi_\cC^E+\Pi_\cC^I)q\,,
    (\Pi_\cC^E+\Pi_\cC^I)q\right\rangle_{{\!\mbox{\tiny $\cM$}}}$ ,
and
\[\left\langle \Pi_\cC^Eq \,, \Pi_\cC^Iq \right\rangle_{{\!\mbox{\tiny $\cM$}}}
= \left\langle \Pi_\cC^Eq \,, (\idty_\cM-\Pi_\cC^E) q \right\rangle_{{\!\mbox{\tiny $\cM$}}}
= \left\langle q\,,\Pi_\cC^E(\idty_\cM-\Pi_\cC^E)q\right\rangle_{{\!\mbox{\tiny $\cM$}}}
=0\,.\]
$J_\cC^E(q)-J_\cD^E(q)=\left\langle (\Pi_\cC^E-\Pi_\cD^E)q \,, q \right\rangle_{{\!\mbox{\tiny $\cM$}}}$
with $\Pi_\cC^E\,\Pi_\cD^E = \Pi_\cD^E$,
and so the formula for the index of positive inertia follows from \eqref{eq:DimDeltaE}.
\hfill $\Box$
\begin{Remark}[Moments of Inertia] \label{rem:J:differences}
Here\\[-6mm]  
\begin{enumerate}[1.]
\item
$J_\cC^E(q)$ equals the moment of inertia of the configuration 
in which all masses of each cluster are joined in its  center of
mass.  By Lemma \ref{lem:CD:comparable} the index of inertia of this quadratic form equals
$d|\cC|$;
\item
$J_\cC^I(q)$ is the sum of the moments of inertia of the clusters,
each referred to the respective center of mass, rather than the
origin;
\item
For $\cC\preccurlyeq\cD$  with $|\cC|=|\cD|+1$, there is a unique cluster $D\in\cD$ which is the disjoint
union $C_1\ \dot{\cup}\ C_2$ of two clusters $C_1,C_2\in \cC$, 
and the other clusters of $\cD$ coincide with the other clusters of $\cC$.
Then 
\begin{align*}
J_{\cC}^E(q)-J_{\cD}^E(q)&= J_{C_1}^E(q)+J_{C_2}^E(q)-J_{D}^E(q)\\
&= m_{C_1} \langle q_{C_1},q_{C_1}\rangle+m_{C_2} \langle q_{C_2},q_{C_2}\rangle -
m_{D} \langle q_{D},q_{D}\rangle\\
&= \frac{m_{C_1}m_{C_2}}{m_{D}}\| q_{C_1}-q_{C_2}\|^2 .
\end{align*}
This measures the squared distance of the barycenters of $C_1$ and $C_2$.
\hfill $\Diamond$
\end{enumerate}
\end{Remark}
Similar statements are true for the external 
kinetic energies, that is, the quadratic forms $K_\cC^E:M^*\to \bR$\quad ($\cC\in \cP(N)$).
\begin{Definition}\phantomsection \label{Graf--Partition}
For $\delta\in(0,1)$ and $k>0$, let 
\[
J^{(k)}:M\to\bR\qmbox{,} J^{(k)}(q)
:= \max \big\{ J_\cC^E(q)+k \delta^{|\cC|} \;\big|\;\cC\in\cP(N)
\big\}\, . 
\]
The \textbf{Graf partition} of the configuration space  $M$ is the
family of subsets 
\begin{equation}
\label{Z4}
\Xi_\cC^{(k)}:=
\left\{q\in M\;\Big|\;J_\cC^E(q)+k\delta^{|\cC|} = J^{(k)}(q) \right\}
\qquad \bigl(\cC\in\cP(N)\bigr).
\end{equation}
\end{Definition}
The dependence on the parameter $k$ is homogeneous: In Minkowski notation
\begin{equation}
\label{homogen}
\Xi_\cC^{(k)}= k^{1/2}\; \Xi_\cC^{(1)}\qquad  \bigl(\cC\in\cP(N), k>0\bigr).
\end{equation}

The Graf partition is a measure theoretic partition of $M$ with respect to Lebesgue measure,
i.e.\ for $\cC\neq \cD$ one has $\lambda^{nd}\big(\Xi_\cC^{(k)}\cap\Xi_\cD^{(k)}\big)=0$.

For small  $\delta\in(0,1)$ (and, by \eqref{homogen}, all $k>0$), the  Graf partition (\ref{Z4})
has the property that for  $\Xi_\cC^{(k)}\cap\Xi_\cD^{(k)}\ne\emptyset$,
the cluster decompositions  $\cC$ and $\cD$ are \textbf{comparable},
i.e., $\cC\preccurlyeq\cD$ or $\cC\succcurlyeq\cD$.
See  \cite[Lemma 12.52]{Kn18} for a proof.

\begin{figure}
\begin{center}
\includegraphics[width=.5\textwidth]{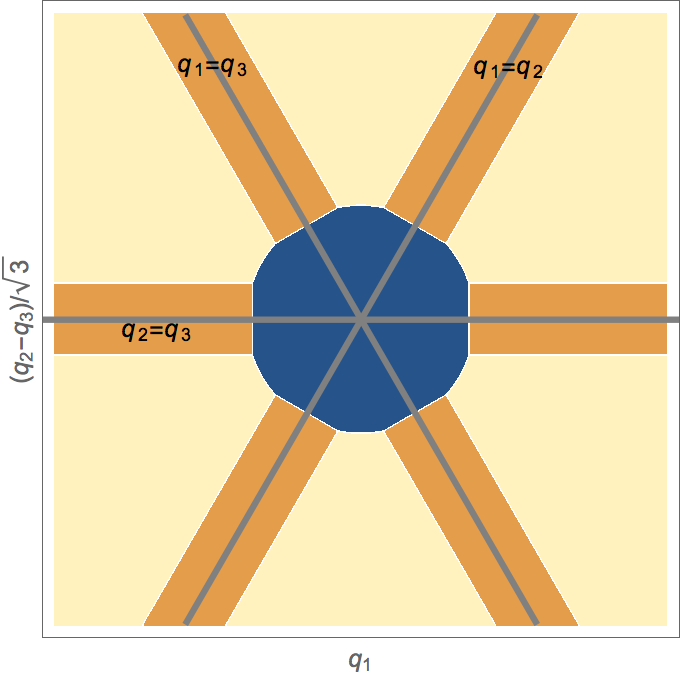}
\end{center}
\vspace*{-5mm}
\caption{Graf partition of the configuration space (center of mass at 0)
for $n=3$ particles in  $d=1$ dimension.
Yellow: $\Xi_{\cC_{\min}}^{(k)}$, Blue: $\Xi_{\cC_{\max}}^{(k)}$; from \cite{Kn18}.}
\phantomsection \label{graf}
\end{figure}

In Figure \ref{graf} we show a Graf partition.

We need quantitative estimates for intracluster and intercluster distances:
\begin{Lemma}  \label{lem:intra:inter}
For small enough $\delta\in(0,\eh)$ in Definition \ref{Graf--Partition} 
there exist constants $C^I,C^E>0$ with $C^I \le C^E/4$ 
such that for all $\cC\in \cP_0(N)$ 
\begin{equation}
\label{quant1}
\|q^I_i\|\le C^I \sqrt{k}\qquad \big(i\in N,\, q\in  \Xi_\cC^{(k)}\big),
\end{equation}
\begin{equation}
\label{quant2}
\|q_{C_i}-q_{C_j}\|\ge C^E \sqrt{k}\qquad \big(C_i\neq C_j\in \cC,\, q\in  \Xi_\cC^{(k)}\big),
\end{equation}
and thus
\begin{equation}
\label{quant3}
\|q_i-q_j\|\ge  \eh \|q_{[i]_\cC}-q_{[j]_\cC}\|\qquad \big([i]_\cC\neq [j]_\cC,\, q\in  \Xi_\cC^{(k)}\big).
\end{equation}
\end{Lemma}
\textbf{Proof:}\quad\\[-6mm]
\begin{enumerate}[$\bullet$]
\item 
To prove \eqref{quant1}, we note that by definition \eqref{Z4} of $\Xi_\cC^{(k)}$ we have 
\[J^I_\cC(q)=J(q)-J^E_\cC(q)=J^E_{\cC_{\min}}(q)-J^E_\cC(q) \le k(\delta^{|\cC|}-\delta^n)
\qquad \big(q\in  \Xi_\cC^{(k)}\big).\]
Now $m_i\|q^I_i\|^2\le J^I_\cC(q)$ and $|\cC|\le n-1$, so that with
$C^I:=(\frac{\delta^{n-1}}{2m_{\min}})^{1/2}$ inequality \eqref{quant1} follows.
\item 
For \eqref{quant2} we compare $\cC=\{C_1,\ldots,C_\ell\}$ with 
\[\cD := \big\{ C_r \mid r\in\{1,\ldots,\ell\}\setminus\{i,j\} \big\} \ \dot{\cup}\  \{D\}\qmbox{, with}D:=C_i\cup C_j.\]
So $\cD\succcurlyeq\cC$ with $|\cD|=|\cC|-1\le n-2$. By Remark \ref{rem:J:differences}.3
\[\frac{m_{C_i}m_{C_j}}{m_{D}}\| q_{C_i}-q_{C_j}\|^2 = J_{\cC}^E(q)-J_{\cD}^E(q)
\ge k(\delta^{|\cD|}-\delta^{|\cC|})\ge k \delta^{n-2}/2.\]
As $m_D=m_{C_i}+m_{C_j}$, $\|q_{C_i}-q_{C_j}\|\ge C^E \sqrt{k}$ with 
$C^E:= (\frac{\delta^{n-2}}{2n\, m_{\max}})^{1/2}$.
\item 
So for $\delta>0$ small, $C^I \le C^E/4$.
Now \eqref{quant3}  follows by the triangle inequality
\[\|q_i-q_j\|\ge  \|q_{[i]_\cC}-q_{[j]_\cC}\| - \|q^I_i\|- \|q^I_j\| 
\ge \|q_{[i]_\cC}-q_{[j]_\cC}\| -{\textstyle \frac {C^E}{2}\sqrt{k}}
\ge \eh\|q_{[i]_\cC}-q_{[j]_\cC}\|.
\hfill\Box\]
\end{enumerate}

The sets
\begin{equation}
\label{Xi:k}
\Xi^{(k)}:=\bigcup_{\cC\in\cP_0(N)} \Xi_\cC^{(k)},
\end{equation}
are neighborhoods of the collision set with $\Delta= \bigcap_{k>0}\Xi^{(k)}$.

Not only is the boundary $\partial\Xi^{(k)}$ of $\Xi^{(k)}\subseteq M$ contained in $\widehat{M}$, 
but there is a lower bound for
$q_{\min}$ defined in \eqref{q:min}:
\begin{Lemma} [Minimal Particle Distance] \label{lem:min:dist}
With $J^{(k)}$ from Definition \ref{Graf--Partition},
\[\partial\Xi^{(k)}
= \{q\in \Xi^{(k)}\mid J^{(k)}(q) = J(q)+k\delta^{n}\}.\]

There is a $C_2>0$ with
\[q_{\min}\ge C_2\sqrt{k}\qquad (q\in \partial\Xi^{(k)}).\]
\end{Lemma}
\textbf{Proof:}
As $J=J^E_{\cC_{\min}}$, $J^{(k)}\ge J+k\delta^n$ by Definition \ref{Graf--Partition}. Thus 
$q\in M$ satisfies $J^{(k)}(q) =J(q)+k\delta^n$ iff $q\in \Xi^{(k)}_{\cC_{\min}}$. 

If additionally $q\in \Xi^{(k)}$, then there is a $\cC\in \cP_0(N)$ with $q\in \Xi^{(k)}_\cC$, too. 
Since $\cC\neq \cC_{\min}$, we conclude that $q\in \partial  \Xi^{(k)}$.

Conversely $\partial\Xi^{(k)} \subseteq  \{q\in \Xi^{(k)}\mid J^{(k)}(q) = J(q)+k\delta^{n}\}$,
since 
\[\partial\Xi^{(k)}= \partial\big(M\setminus \Xi^{(k)}\big) =\partial \big({\rm int}(\Xi^{(k)}_{\cC_{\min}})\big).\]

Let $q\in \partial\Xi^{(k)}\cap \partial\Xi^{(k)}_\cC$ and indices $i,j\in N$ so that $q_{\min}(q)=\|q_i-q_j\|$.
\begin{enumerate}[$\bullet$]
\item 
If $[i]_\cC \neq [j]_\cC$, then it follows from \eqref{quant3} and \eqref{quant2} that
$\|q_i-q_j\|\ge C^E\sqrt{k}/2$.
\item 
Otherwise $[i]_\cC = [j]_\cC$, but $q\in \Xi^{(k)}_{\cC_{\min}}$ so that $[i]_{\cC_{\min}} \neq [j]_{\cC_{\min}}$.
For $q\in \partial \Xi^{(k)}$ in particular $J^I_{\{i,j\}} \geq k (\delta^{n-1}-\delta^n)$ for all $i \neq j$, so that
$q_{\min} \geq C k^{1/2}$.
\hfill $\Box$\\[2mm]
\end{enumerate}

Later, in \eqref{F:m}, we will define a sequence of hypersurfaces in configuration space $\hM$ to which
our Poincar\'e surfaces in $\hP$ are to project.
Therefore we now estimate the Riemannian hypersurface volumes of $\partial\Xi^{(k)}$, 
intersected with balls 
\[B_R:=\{q\in M\mid \|q\|_\cM\le R\},\] 
whose radius $R$ goes to infinity as $k\searrow 0$. 
In the Euclidean space $(M,\langle \cdot,\cdot \rangle_M )$ (see \eqref{inner:prod:M}), 
the $(nd-1)$-dimensional Riemannian hypersurface volume element is denoted by $d\cF$.
To obtain an easy upper bound, we instead estimate the $d\cF$-volumes of the cylinders
\begin{equation}
\label{Z:cyl}
Z_\cC^{(k)} :=\{q\in M\mid J_\cC^I(q)=\eta_\cC\}\cong 
\Delta^E_{\cC}\times \{q\in \Delta^I_{\cC}\mid J_\cC^I(q)=\eta_\cC\} \quad\ (\cC\in \cP_0(N)),
\end{equation}
intersected with $B_R$, with $\eta_\cC:=k (\delta^{|\cC|}-\delta^n)>0$.
$Z_\cC^{(k)}$ is diffeomorphic to $\bR^{d|\cC|}\times S^{d(n-|\cC|)-1}$.
Notice that, unlike on $\partial  \Xi_\cC^{(k)}$, the potential $V$ may diverge on $Z_\cC^{(k)}$
and is even undefined on the measure zero set $Z_\cC^{(k)}\cap \Delta$.
\begin{Lemma}  \label{lem:F:area}
There is a decomposition of the boundary $\partial \Xi^{(k)}$ as the union of
\begin{equation}
\label{deco}
\partial \Xi_\cC^{(k)}\cap \partial \Xi^{(k)} \subseteq Z_\cC^{(k)}\qquad(\cC\in \cP_0(N)).
\end{equation}
There exists $C>0$ with
\[ \int_{\partial\Xi^{(k)}}\idty_{B_R}\,d\cF\leq C\, k^{(d-1)/2}\, R^{d(n-1)} \qquad(0<k\le 1 \le R).\]
\end{Lemma}
\textbf{Proof:}
By definition,
$\Xi^{(k)}=\{q\in M\mid J^{(k)}_0(q)=J^{(k)}(q) \}$ with
\[J^{(k)}_0:M\to\bR\qmbox{,} J^{(k)}_0(q)
:= \max \big\{ J_\cC^E(q)+k \delta^{|\cC|} \;\big|\;\cC\in\cP_0(N)\big\}.\]
So with $\eta_\cC=k (\delta^{|\cC|}-\delta^n)$, using \eqref{Z8},
\[\partial \Xi^{(k)}= \{q\in M\mid \exists \,\cC\in \cP_0(N):  J_\cC^I(q)=\eta_\cC,\,
\forall \cD\in\cP_0(N): J_\cD^I(q) \ge \eta_{\cD} \}.\]
On the other hand, by \eqref{Z4},
\[\partial \Xi_\cC^{(k)}\subseteq
\left\{q\in M\;\Big|\;J_\cC^I(q)\le\eta_\cC \right\}
\qquad \bigl(\cC\in\cP_0(N)\bigr).\]
Thus $J_\cC^I(q) = \eta_\cC$ for $q\in \partial \Xi_\cC^{(k)}\cap \partial \Xi^{(k)}$, showing \eqref{deco}. 
This implies
\[\int_{\partial\Xi^{(k)}}\idty_{B_R}\,d\cF
= \sum_{\cC\in \cP_0(N)} \int_{\partial \Xi_\cC^{(k)}\cap\partial\Xi^{(k)}}\idty_{B_R}\,d\cF 
\leq  \sum_{\cC\in \cP_0(N)}\int_{Z_\cC^{(k)}} \idty_{B_R}\,d\cF\]
with the cylinders $Z_\cC^{(k)}$. 
So by \eqref{eq:DimDeltaE} and \eqref{eq:DimDeltaI}
\[\int_{Z_\cC^{(k)}} \idty_{B_R}\,d\cF\le 
v_{d|\cC|}\, s_{d(n-|\cC|)-1}\,(k(\delta^{|\cC|}-\delta^n))^{(d(n-|\cC|)-1)/2}\, R^{d|\cC|},\]
with the volume $v_m$ of the $m$-dimensional unit ball and the surface area $s_m$ of the sphere $S^m$.
The estimate follows, since $\max\{|\cC|\mid \cC\in\cP_0(N)\}=n-1$.
\hfill $\Box$\\[2mm]

From Lemma \ref{lem:F:area} one concludes that $\lim_{k\searrow 0} \int_{\partial\Xi^{(k)}}\idty_{B_R}\,d\cF=0$,
provided that $R\equiv R(k)=o\big(\! k^{-x_{\max}}\! \big)$  with $x_{\max}:=\frac{d-1}{2d(n-1)}$. Thus our assumption 
$d\ge 2$ allows for divergence of~$R$.

Accordingly, if we set for $x\in(0,x_{\max})$
\begin{equation}
\label{fixing:parameters}
k(m):= 4^{-m}\mbox{ and }R(m):=4^{mx}\qquad(m\in \bN)
\end{equation}
in
\begin{equation}
\label{F:m}
\cF_m:=\bigcup_{\cC\in \cP_0(N)} \cF_{m,\cC}\qmbox{with} 
\cF_{m,\cC}:= \partial \Xi^{(k(m))}\cap \Xi_\cC^{(k(m))}\cap B_{R(m)},
\end{equation}
then $\int_{\cF_m}d\cF=\cO(2^{-(d-1)(1-x/x_{\max})m})\stackrel{m\to\infty}{\longrightarrow}0$.\\
When we are to include integration over momenta, we will have to restrict $x>0$ further, see the proof of
Proposition \ref{prop:area}.

The significance of that family $(\cF_m)_{m\in\bN}$ of hypersurfaces is clarified by the following lemma:

\begin{Lemma}  \label{lem:aa:Fm}
The forward configuration space trajectory $t\mapsto q(t,x)$ of any initial condition $x\in\Coll$
intersects all but finitely many hypersurfaces  ${\cF_m}$.
\end{Lemma}
\textbf{Proof:}
The trajectory has to enter all neighborhoods $\Xi^{(k(m))}$, see \eqref{Xi:k}, of the collision set $\Delta$.
On the other hand, by definition \eqref{eq:Coll} of $\Coll$,  the limit 
$\lim_{t \nearrow T(x)} q(t,x)\in M$ exists.
So for the positive time interval $[0,T(x))$ the trajectory stays in a bounded region of $\hM$, and, by
\eqref{fixing:parameters},  is contained in $B_{R(m)}$ for all $m\ge m_0$. The claim follows from Definition \eqref{F:m}.
\hfill $\Box$
\begin{Remark}[Symmetries of Hypersurfaces] \quad\\
Later, when we estimate the symplectic volumes of the Poincar\'{e} surfaces erected over the hypersurfaces 
$\cF_{m,\cC}\subseteq M$, we will break down
that high-dimensional integration. One basic step is the factorization $M=\Delta^E_\cC\oplus\Delta^I_\cC$,
and its sub-factorizations, see \eqref{M:E:I}. The cylinder $Z_\cC^{(k)}$ defined in \eqref{Z:cyl}
respects these:
\begin{equation}
\label{Z:Delta:S}
Z_\cC^{(k(m))} = \Delta^E_\cC\times S_{m,\cC} \qmbox{with}
S_{m,\cC}:=\{q\in \Delta^I_\cC\mid J^I_\cC(q)=k(m)(\delta^{|\cC|}-\delta^n)\}. 
\end{equation}
So $S_{m,\cC}$ is a sphere of dimension $d (n - | \cC | )-1$, and by \eqref{deco} and \eqref{F:m},
\[\cF_{m,\cC}\subseteq\Delta^E_\cC\times S_{m,\cC}.\]
A decomposition of the factor $\Delta^E_\cC$, adapted to the potential $V$, will be performed using Jacobi
coordinates.
\hfill $\Diamond$
\end{Remark}
\paragraph{Jacobi Coordinates and Maximal Chains}\quad\\
We will now refine even further the decomposition \eqref{F:m} of  $\cF_{m}$ into the $\cF_{m,\cC}$.
The reason is that we have to cope with the following problem.
One could expect that the configuration space trajectory $t\mapsto q(t,x)$
of initial condition $x\in\Coll$, finally colliding in the set partition $\cC:={\rm SP}(x)\in \cP_0(N)$,
should intersect the hypersurfaces $\cF_{m,\cC}$, for large enough $m\in \bN$. 

However, this need not be the case, since some tight subcluster of particles could form before collision takes
place. In Figure \ref{graf} that would correspond to a trajectory entering the region $\Xi_{\cC_{\max}}^{(k)}$
through one of the channels. This then could lead to intersections of the trajectory with hypersurfaces 
$\cF_{m,\cD}$, with $\cD\neq \cC$ for all large~$m$.

At least we can assure that this can only occur if $\cD\preccurlyeq\cC$:
\begin{Lemma}[From Finer to Coarser Partitions]\label{lem:finer:coarser}\quad\\  
Let $x\in\Coll$, $\cD\in \cP_0(N)$ and $(m_i)_{i\in\bN}$, $(t_i)_{i\in\bN}$ be strictly increasing sequences with 
$q(t_i,x)\in \cF_{m_i,\cD}$. Then $\cD\preccurlyeq{\rm SP}(x)$.
\end{Lemma}
\textbf{Proof:}
As $\lim_{i\to\infty} k(m_i)=0$, $\lim_{i\to\infty}q(t_i,x)\in \Delta_\cD^E$ so that 
$\lim_{i\to\infty} t_i=T(x)$. 
From \eqref{Xi:Null} it follows that $\Delta_\cD^E=\dot{\bigcup}_{\cC\succcurlyeq \cD} \Xi_\cC^{(0)}$. So
$\cD\preccurlyeq{\rm SP}(x)$.
\hfill $\Box$\\[2mm]
The external momentum $p^E_\cC(t,x)$ has a limit as $t\nearrow T(x)$ {\em if}
$\cC={\rm SP}(x)$. This follows simply, since then for $\cC=(C_1,\ldots ,C_k)$ by definition the 
particles of the same $C_i$ converge to the same point, and these points are different for $C_i$, $C_j$
with $i\neq j$.

However, the external momentum $p^E_\cD(t,x)$ may diverge in the limit $t\nearrow T(x)$, if
${\rm SP}(x)\succneqq\cD$. Thus when in Section \ref{sec:Definition:P} we erect the Poincar\'{e} surfaces
$\cH_m$ in the energy shell over the $\cF_m$, we have to make them large enough so that
they are hit in spite of this divergence.
On the other hand the symplectic volume of the Poincar\'{e} surface should go to zero as $m\to \infty$.

In order to find a definition of the $\cH_m$ meeting these requirements, 
we (measure theoretically) decompose the $\cF_{m,\cC}$ into subsets,  indexed by maximal chains
\begin{equation}
\label{def:MC}
 \cC_1\succneqq\cdots \succneqq\cC_k\qquad\mbox{from } 
\cC_1:= \cC_{\max}\mbox{ to }\cC_k:=\cC\mbox{, with }|\cC_\ell|=\ell.
\end{equation}
We denote the set of maximal chains ending at $\cC$ by ${\rm MC}(\cC)$.

The maximal chain induces a variant of Jacobi coordinates,
not for the positions of the bodies but for the cluster barycenters of $\cC$.
To define them, we use a double index for the clusters of the set partitions: 
\[\cC_\ell=\{C_{\ell,1},\ldots, C_{\ell,\ell} \}\qquad (\ell=1,\ldots,k=|\cC|).\]
By Remark \ref{rem:J:differences}.3 there are uniquely two indices $1\le L_\ell<R_\ell\le \ell$ and an index $1\le U_\ell\le\ell-1$ with
\begin{equation}
\label{block:merge}
C_{\ell-1,U_\ell} = C_{\ell,L_\ell}\ \dot{\cup}\ C_{\ell,R_\ell}\qquad(\ell=2,\ldots,k),
\end{equation}
whereas all other blocks  $C_{\ell,i}\in \cC_\ell$ equal blocks $C_{\ell-1,\pi_\ell(i)}\in \cC_{\ell-1}$.
This attributes to the maximal chain $K=(\cC_1,\ldots, \cC_k)\in {\rm MC}(\cC)$ the linear isomorphism
\begin{equation}
\label{Jacobi}
Q\equiv Q_K=(Q_{K,1},\ldots,Q_{K,k}):\Delta^E_{\cC}\longrightarrow \bigoplus_{j=1}^{k} \bR^d
\end{equation}
with the {\em Jacobi coordinates}
\[Q_1:=q_{N}\equiv q_{C_{1,1}}\qmbox{and} 
Q_\ell := q_{C_{\ell, L(\ell)}}  -  q_{C_{\ell, R(\ell)}} \qquad (\ell=2,\ldots,k).\]
So $Q_1$ is the center of mass of all particles, and the other $Q_\ell$ are the differences of
the barycenters of the two clusters to be merged.

The external configuration space region that we attribute to the maximal chain 
$K =(\cC_1,\ldots, \cC_k)\in {\rm MC}(\cC)$ is its {\em Jacobi space}
\begin{equation}
\label{Jacobi:space}
\cJ_K:=\big\{q\in \Delta^E_\cC \mid \forall \ell\in \{2,\ldots,k\}:
\, \|Q_{K,\ell} \|= \!\!\!\!\min_{\cD\in\cP^{(\ell-1)}, \cD\succcurlyeq \cC_{\ell}}
\| q_{C_{\ell, \tilde{L}(\ell)}}  -  q_{C_{\ell, \tilde{R}(\ell)}} \|\big\}.
\end{equation}
Here $\tilde{L}(\ell)$ and $\tilde{R}(\ell)$ index the clusters of $\cC_\ell$ to be merged in $\cD$.

For $n=1$ and $n=2$ particles there is only one maximal chain $K$, and $\cJ_K=\Delta^E_\cC$. For all $n$
\[\Delta^E_\cC = \bigcup_{K\in {\rm MC}(\cC)}\cJ_K\qquad(\cC\in \cP_0(N)),\]
since $\cP(N)$ is a lattice.
Conversely, the $\cJ_K$ are disjoint w.r.t.\ Lebesgue measure, 
\[\lambda^{d|\cC|}(\cJ_K\cap \cJ_L)=0\qquad(K\neq L\in {\rm MC}(\cC)),\] 
since $\|Q_{K,\ell}\| = \|Q_{L,\ell}\|$ for $q\in \cJ_K\cap \cJ_L$, but $Q_{K,\ell}\neq Q_{L,\ell}$ for some $\ell$.

\begin{figure}
\begin{center}
\includegraphics[width=.7\textwidth]{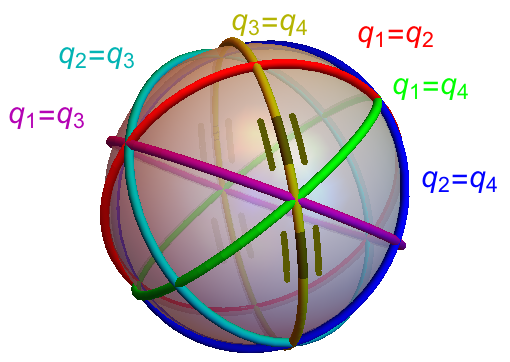}
\end{center}
\vspace*{-5mm}
\caption{Collision subspaces for $n=4$ particles in  $d=1$ dimension,
with barycenter at $0\in \bR^4$. Shown are only the intersections of these hyperplanes $\{q_i=q_j\}$ 
with the sphere $S^2$. \newline
For the maximal chain $K=\{\cC_1,\cC_2,\cC_3\}$ with 
$\cC=\cC_3=\{\{1\},\{2\},\{3,4\}\}$, $\cC_2=\{\{2\},\{1,3,4\}\}$ and $\cC_1=\{\{1,2,3,4\}\}=\cC_{\max}$, 
the Jacobi space  $\cJ_K\subseteq\Delta^E_\cC$ (that is, in $\{q_3=q_4\}$) 
and a hypersurface $\cF_{m,K}$ appear in darker color.}
\phantomsection \label{sphereColl}
\end{figure}

Finally, this induces a decomposition of the hypersurfaces \eqref{F:m}, given by
\begin{equation}
\label{F:mK}
\cF_{m,\cC}=\bigcup_{K\in {\rm MC}(\cC)}\cF_{m,K}\qmbox{with}
\cF_{m,K}:=\cF_{m,\cC}\cap(\cJ_K\times \Delta^I_\cC).
\end{equation}
The construction is shown in Figure \ref{sphereColl}.
We can locally dominate the inter-cluster potential, using this decomposition:
\begin{Lemma}  \label{lem:JJ:Q}
For all $\cC\in \cP_0(N)$ and maximal chains $K=(\cC_1,\ldots, \cC_k)\in {\rm MC}(\cC)$
\[ J_{\cC_1}^E(q)=m_{N}\|Q_1\|^2\]
and
\[J_{\cC_{\ell}}^E(q)-J_{\cC_{\ell-1}}^E(q)=
\frac{ m_{C_{\ell, L(\ell)}}  m_{C_{\ell, R(\ell)}} } 
{m_{C_{\ell, L(\ell)}} + m_{C_{\ell, R(\ell)}}} \|Q_\ell\|^2\qquad (\ell=2,\ldots,k), \]
with $k=|\cC|$. Furthermore, $V=V_\cC^E+V_\cC^I$ with {\bf intra-} respectively {\bf inter-cluster potential} 
\begin{equation}
\label{V:E:I}
V_\cC^E(q) := \sum_{i \nsim j, i<j} V_{i,j}(q_i - q_j) \mbox{ resp. } \ 
V_\cC^I(q) := \sum_{i \sim j, i<j} V_{i,j}(q_i - q_j)\qquad(q\in \hM),
\end{equation}
and for suitable $C_I, C_{II}>0$
\begin{align}
\label{V:ineq}
\left| V_\cC^E(q) \right| 
&\le C_I +C_{II}  \sum_{i,j:[i]_\cC\neq[j]_\cC } \|q_i-q_j\|^{-\alpha}\nonumber\\
&=C_I+\cO\big(\|Q_k\|^{-\alpha}\big)
\quad \big(q\in \Xi_{\cC}^{(k)}\cap (\cJ_K\times \Delta_\cC^I) \setminus\Delta\big).
\end{align}
\end{Lemma}
\textbf{Proof:}\\
As $\cC_1=\cC_{\max}=\{N\}$, the formula for $J_{\cC_1}^E$ follows directly from Definition \eqref{Z8}.

By Remark \ref{rem:J:differences}.3
\[
J_{\cC_{\ell}}^E(q)-J_{\cC_{\ell-1}}^E(q)
= \frac{m_{C_{\ell, L(\ell)}}  m_{C_{\ell, R(\ell)}}}{m_{C_{\ell-1,U_\ell}}} 
\| q_{C_{\ell,L(\ell)}}-q_{C_{\ell,R(\ell)}} \|^2,
\]
and $m_{C_{\ell-1,U_\ell}} =m_{C_{\ell, L(\ell)}} + m_{C_{\ell, R(\ell)}}$.

By admissibility, the first inequality of \eqref{V:ineq} is true:
\begin{itemize}
\item 
Definition \ref{def:admissible} states that $\,\lim_{\|q\|\to\infty}V_{i,j}(q)=0$. So for some 
$C'_I>0$ we have $|V_{i,j}(q)| \le C'_I$ if $\|q\|\ge 1$.
\item 
Integrating \eqref{D2V} twice along the line between $q$ and $q/\|q\|$, we obtain for some 
$C'_{II}>0$
\[|V_{i,j}(q)|\le C'_{II}\|q\|^{-\alpha}\qquad(\|q\|\le 1).\]
\end{itemize}
We set $C_I := {n\choose 2}C'_I$ and $C_{II} := {n\choose 2}C'_{II}$.

By \eqref{quant3} of Lemma \ref{lem:intra:inter} 
\[\sum_{i,j:[i]_\cC\neq[j]_\cC } \|q_i-q_j\|^{-\alpha} 
\le  c_1\sum_{C,D\in \cC:\, C\neq D}\|q_{C}-q_{D}\|^{-\alpha}
\qquad \big(q\in \Xi^{(k)}_\cC\setminus\Delta\big)\]
with $c_1:={n\choose 2}2^{\alpha}$.
The definition \eqref{Jacobi:space} of Jacobi space $\cJ_K$ then implies that $\|q_{C}-q_{D}\|\ge \|Q_k\|$,
leading to the order estimate in \eqref{V:ineq}.
\hfill $\Box$\\[2mm]
%

%
\section{Definition and Symplectic Volume of the Poin\-car\'{e} Surfaces}
\label{sec:Definition:P}
%
%
Within an energy surface $\Sigma_E$ we will define in \eqref{H:m} a family  $(\cH_m)_{m\in \bN}$ 
of hypersurfaces.
Any collision orbit in $\Sigma_E$ must intersect almost all of the $\cH_m$.
We will also estimate their symplectic volumes, in Proposition \ref{prop:area}. 
To perform that task, we will use adapted coordinates. As the coordinate changes are symplectomorphisms,
these preserve the volume.
We begin by presenting the first such coordinate change, indexed by a cluster decomposition $\cC$.

Using the natural identification $T^* M \cong M \times M^*$, the decomposition \eqref{M:E:I} 
of $M$ expands to a $T^*M$-orthogonal decomposition
\begin{equation}
\label{TStarM:decomp}
T^* M \ =\ T^* \Delta_\cC^E \oplus T^* \Delta_\cC^I \ =\ 
T^* \Delta_\cC^E \oplus \bigoplus_{C\in\cC} T^* ( \Delta_C^I)
\end{equation}
of phase space. The projections onto the corresponding components are denoted by 
$\widehat{\Pi}_\cC^E$ respectively $\widehat{\Pi}_\cC^I$.

\begin{Remark} 
[Understanding the Projections $\widehat{\Pi}_\cC^E$ and $\widehat{\Pi}_\cC^I$]\quad\\ 
For a subspace $N\subseteq M$ of an $\bR$-vector space $M$ there is no natural embedding 
$T^*N\subseteq T^*M$, although naturally $TN\subseteq TM$. So \eqref{TStarM:decomp}
necessitates a definition.

Here, using the musical isomorphism 
\[I:M\to M^*\qmbox{,} v\mapsto \langle v,\cdot\rangle_M=\langle\cM v,\cdot\rangle,\] 
we have
\[\langle I(v),I(w)\rangle_{M^*} = \langle\cM v, \cM w\rangle_{M^*} = \langle v, \cM w\rangle
= \langle v, w\rangle_M \qquad (v,w\in M).\]
So when we identify $T^* \Delta_\cC^E$ with $\Delta_\cC^E\times I(\Delta_\cC^E)\subseteq T^*M$
and similarly $T^* \Delta_\cC^I$ with $\Delta_\cC^I\times I(\Delta_\cC^I)\subseteq T^*M$, these
subspaces are $\langle \cdot,\cdot\rangle_{T^*M}$-orthogonal, see \eqref{TsM:product}.
\hfill $\Diamond$
\end{Remark}
By this, we indeed did define symplectic coordinates:

\begin{Lemma}
\label{lem:ClusterSymCoord}
For all $\cC \in \cP( N)$, the vector space isomorphism
\begin{equation*}
\left( \widehat{\Pi}_\cC^E, \widehat{\Pi}_\cC^I \right): T^*M \rightarrow 
T^* \Delta_\cC^E \oplus \bigoplus_{\bfC\in\cC} T^* ( \Delta_\bfC^I)
\end{equation*}
is symplectic w.r.t. to the canonical symplectic forms on the respective cotangent bundles. \hfill $\Box$
\end{Lemma}
Similarly, for all $\cC\in \cP_0(N)$ and maximal chains $K=(\cC_1,\ldots, \cC_k)\in {\rm MC}(\cC)$
 the {\em Jacobi map} ${\rm JM}_K$ of phase spaces is symplectic:
\begin{Lemma}[Jacobi Map]\label{lem:JM}  \quad\\ 
The inverse cotangent lift of the Jacobi transformation \eqref{Jacobi} has the form
\[ {\rm JM}_K:= (Q_K^*)^{-1}: T^*\Delta^E_\cC\to T^*(\oplus_{j=1}^{k} \bR^d)\qmbox{,}\ 
(q^E_\cC,p^E_\cC)\mapsto (Q,P)\]
with $(Q,P)=(Q_1,\ldots,Q_k\,,\,P_1,\ldots,P_k)$,
\[Q_1=q_{N}\equiv q_{C_{1,1}}\qmbox{,} 
Q_\ell = q_{C_{\ell, L_\ell}}  -  q_{C_{\ell, R_\ell}} \qquad (\ell=2,\ldots,k)\]
with the notation \eqref{block:merge}, and
\[P_1=p_N\equiv p_{C_{1,1}}\qmbox{,}
P_\ell = \frac{ m_{C_{\ell, R_\ell}} p_{C_{\ell, L_\ell}} - m_{C_{\ell, L_\ell}} p_{C_{\ell, R_\ell}} }
{m_{C_{\ell, L_\ell}}+m_{C_{\ell, R_\ell}} } 
 \qquad (\ell=2,\ldots,k).\]
\end{Lemma}
\textbf{Proof:}
This follows from the facts that 
\begin{enumerate}[$\bullet$]
\item 
for $C\subseteq N$, $C\neq \emptyset$ 
the same spatial components of the $\bR^d$-valued functions $q_C$ and $p_C$ have Poisson bracket one,
\item 
whereas different components of $q_C$ and $p_C$ have Poisson bracket zero,
\item 
that for $i\in N\setminus C$ the components of $q_i$ have Poisson bracket zero with the ones of $p_C$,
\item 
and that $q_{C_{\ell-1, U_\ell}}
= (m_{C_{\ell, L_\ell}}q_{C_{\ell, L_\ell}}  + m_{C_{\ell, R_\ell}} q_{C_{\ell, R_\ell}})/
m_{C_{\ell-1, U_\ell}}$.\hfill $\Box$\\[2mm]
\end{enumerate}

A straightforward calculation shows that kinetic and total energy split into their internal and external parts:

\begin{Lemma}[External and Internal Energies]
\label{pro:BaryRelEnergy}\quad\\[-6mm]
\begin{enumerate}[$\bullet$]
\item 
If we define \emph{barycentric} and \emph{relative kinetic energy} by 
\[K^E_\cC := K \circ \widehat{\Pi}^E_\cC\qmbox{,}K^E_\cC(q)=\sum_{C\in\bC}K^E_C(q)\mbox{ with }
K^E_C(q)=\frac{\|p_C\|^2}{2m_C}\]
respectively $K^I_\cC := K \circ \widehat{\Pi}^I_\cC$, 
we get $K = K^E_\cC + K^I_\cC$ for all $\cC \in \cP( N)$.
\item 
Hence, $H = H^E_\cC + H^I_\cC$, with $H_\cC^E = K_\cC^E + V_\cC^E$ resp. 
$H_\cC^I = K_\cC^I + V_\cC^I$, see \eqref{V:E:I}
\end{enumerate} 
\end{Lemma}
In general, however, $H^E_\cC$ and $H^I_\cC$ only Poisson-commute for $\cC=\cC_{\max}$.

Next we erect Poincar\'{e} sections $\cH_m\subseteq \Sigma_E$ over $\cF_m$: 
\begin{equation}
\label{H:m}
\cH_m:=\bigcup_{\cC\in \cP_0(N)} \cH_{m,\cC}
\qmbox{with}
\cH_{m,\cC}:=\bigcup_{K\in {\rm MC}(\cC)}\cH_{m,K} 
\end{equation}
and, using the Jacobi map ${\rm JM}_K$ of Lemma \ref{lem:JM} and Definition \eqref{F:mK} of $\cF_{m,K}$, 
\begin{align}
\label{H:mK}\nonumber
\cH_{m,K} :=
 \Big\{(q,p)\in \Sigma_E\ \Big|\ & q \in \cF_{m,K} , \|P_1 \|^2\le 4^{\beta m},\\
& \forall \ell\in\{2,\ldots,k\}: \|P_\ell \|^2\le 4^{\beta m}\|Q_\ell \|^{-\alpha}
\Big\}. 
\end{align}
\begin{Remark}[The Poincar\'{e} Sections] \quad\\[-6mm]
\begin{enumerate}[1.]
\item 
The $\cH_{m,K}$ are well-defined, since for all $q \in \cF_{m,\cC}$ the $Q_\ell$ ($\ell\in\{2,\ldots,k\}$) 
are nonzero, using Lemma \ref{lem:JJ:Q}:
\[J_\cC^E(q)-J_\cD^E(q)\ge 4^{-m}\big(\delta^{|\cD|}-\delta^{|\cC|}\big)>0\qquad (\cD\succneqq \cC).\]
\item 
Notice that the kinetic energy $K_{\cC_{\max}}^E=\|P_1 \|^2/(2m_N)$ 
of the center of mass is a constant of the motion for $\Phi$.
\item 
For regular values $E$ of $V$ the $\cH_{m,\cC}$ are codimension one $\partial$-submanifolds of 
the submanifold $\Sigma_E\subseteq \hP$.
For critical values $E$ of $V$ we could redefine $\hM$ by omitting all critical points $q\in \hM$.
Since they are covered by rest points $(q,0)\in \Sigma_E$, this does not change the set
$\Coll_E\subseteq \Sigma_E$, but guarantees the submanifold property of the $\cH_{m,\cC}$.

However, this redefinition of $\hM$ is not necessary, as we will work with the subsets 
$\cH_{m,K}^\pm\subseteq \cH_{m,K}$,
see \eqref{H:m:K:pm} below, that are by their definition $\partial$-submanifolds of $\hP$, see Lemma 
\ref{lem:pa:submani} below.
\hfill $\Diamond$
\end{enumerate}
\end{Remark}

The cylinders $Z_\cC^{(k)}$ are hypersurfaces of $\hM$, thus oriented by an orientation of their normal
bundle. This is spanned and oriented by the unit vector field
\[N_\cC: Z_\cC^{(k)}\to T_{Z_\cC^{(k)}}M\qmbox{,} q\mapsto \Big(q,\frac{q_\cC^I}{\| q_\cC^I\| }\Big).\]
This allows to decompose the Poincar\'{e} sections in two mirror symmetric parts (and a remaining set of 
zero $\Omega_{nd-1}$-volume), setting 
\begin{equation}
\label{H:m:K:pm}
\cH_{m,K}^\pm:=\{(q,p)\in \cH_{m,K}\mid \pm p(N_\cC(q)) > 0\}\qquad (K\in {\rm MC}(\cC)).
\end{equation}
Correspondingly, 
\begin{equation}
\label{H:m:pm}
\cH_m^\pm:=\bigcup_{\cC\in \cP_0(N)} \cH_{m,\cC}^\pm
\qmbox{with}
\cH_{m,\cC}^\pm:=\bigcup_{K\in {\rm MC}(\cC)}\cH_{m,K}^\pm. 
\end{equation}
The $\cH_{m,K}^-$ consist of points on trajectories entering the neighborhood $\Xi^{(k(m))}$ 
of the collision set. Their embeddings are denoted by
\begin{equation}
\label{i:m}
\imath_{m,K}^\pm:\cH_{m,K}^\pm \to T^*\hM.
\end{equation}
\begin{Lemma}  \label{lem:pa:submani}
The hypersurfaces $\cH_{m,K}^\pm\subseteq \Sigma_E$ $(m\in \bN)$ are transverse to the flow.
They are symplectic $\partial$-manifolds. Thus (see \eqref{eq:DefOmega}) 
\begin{equation}
\label{Omega}
\Omega:=\Omega_{nd-1}
\end{equation}
induces volume forms 
$(\imath_{m,K}^\pm)^*\Omega$ on the  $\cH_{m,K}^\pm$, and their volumes are finite.
\end{Lemma}
\textbf{Proof:}
The $\cH_{m,K}^\pm$ are transverse to the vector field $X_H$ generating the flow $\Phi$, 
since they project to the hypersurface $\cF_{m,K}$, $X_H(q,p)=(\cM^{-1}p,-\nabla V(q))$
and $\langle \cM^{-1}p,N_\cC(q)\rangle_M=p(N_\cC(q))\neq0$.

Since $p\neq0$ in \eqref{H:m:K:pm}, the $(q,p)\in \cH_{m,K}^\pm$ are regular points of the Hamiltonian $H$.
The transversality property w.r.t.\ $X_H$  shows that $\cH_{m,K}^\pm$ are $\partial$-manifolds of $\hP$.

It is a standard argument, see {\em e.g.} {\sc McDuff} and {\sc Salamon} \cite[Lemma 8.2]{MS}, 
that from such a transversality
it follows that the submanifold (or $\partial$-manifold) is symplectic.

As the closures $\overline{\cH_{m,K}^\pm}$ are compact, they have finite symplectic $\Omega$-volume. 
\hfill $\Box$\\[2mm]
In order to estimate that volume, we project it to $T^* \cF_{m,K}$, by
\begin{equation}
\label{def:pro:f}
n_{m,K}^\pm: \cH_{m,K}^\pm \to T^* \cF_{m,K}\quad\mbox{,}\quad
(q,p)\mapsto \big(q,p-p(N_\cC(q)) N_\cC^{\flat}(q)\big).
\end{equation}
The cotangent bundle $T^* \cF_{m,K}$ carries the canonical symplectic form $\omega_\cF$.
Similar to \eqref{eq:DefOmega} and \eqref{Omega} we denote the symplectic volume on $T^* \cF_{m,K}$ by 
\[\widetilde{\Omega} := \frac{(-1)^{\lfloor k/2 \rfloor}}{k!}\omega_\cF^{\wedge k}, \qquad\mbox{with } k:=nd-1.\]
With respect to the embeddings \eqref{i:m} one has 
\begin{equation}
\label{eq:symp:symp}
(\imath_{m,K}^\pm)^*\omega_0 = (n_{m,K}^\pm)^*\omega_\cF,
\end{equation}
see Theorem C of \cite{Fleischer_Knauf_2018}. 

By reversibility the two Poincar\'{e} sections $\cH_{m,K}^-$ and $\cH_{m,K}^+$ have a common image 
\begin{equation}
\label{tilde:H:mk}
\widetilde{\cH}_{m,K}:=n_{m,\cC}^\pm(\cH_{m,K}^\pm)\subseteq  T^* \cF_{m,K}.
\end{equation}
Equation \eqref{eq:symp:symp} allows to estimate the $\widetilde{\Omega}$--volume 
of $\widetilde{\cH}_{m,K}$ instead of the $\Omega$\,--volume of $\cH_{m,K}^\pm$.
To do this, we use the symplectic Jacobi map ${\rm JM}_K$ of Lemma \ref{lem:JM} to present 
$\widetilde{\Omega}$ in a form adapted to $\widetilde{\cH}_{m,K}$.
\begin{Proposition}  \label{prop:area}\quad\\
By choosing the constants $x>0$ in \eqref{fixing:parameters} and  $\beta>0$ in \eqref{H:mK} 
small enough,
\[\lim_{m\to\infty}\int_{\cH_{m,K}^\pm}\Omega =0\qquad\big(\cC\in\cP_0(N),\ K\in  {\rm MC}(\cC)\big)\]
for $d\ge2$ dimensions. So then $\lim_{m\to\infty}  \int_{\cH_{m}^\pm}\Omega =0$.
\end{Proposition}
\textbf{Proof:}
\begin{enumerate}[1.]
\item 
By \eqref{eq:symp:symp} and \eqref{tilde:H:mk}, 
$\int_{\cH_{m,K}^\pm}\Omega = \int_{\widetilde{\cH}_{m,K}}\widetilde{\Omega}$.
\item 
Since $q_{\min}(q)\ge C_2\sqrt{k}$
for $q\in \partial \Xi^{(k)}$ (Lemma \ref{lem:min:dist}), and by \eqref{deco},
\[\widetilde{\cH}_{m,K}\subseteq \widehat{\cH}_{m,K}\qmbox{and thus} 
\int_{\widetilde{\cH}_{m,K}}\widetilde{\Omega} \le  \int_{\widehat{\cH}_{m,K}}\widetilde{\Omega}\]
with, see \eqref{Z:Delta:S}, 
\begin{align*}
\widehat{\cH}_{m,K}:=& \Big\{(q,p)\in T^*\big((\Delta^E_\cC\cap B_{R(m)})\times S_{m,\cC}\big)
\ \Big| \
\|P_1 \|^2\le 4^{\beta m},\\ 
&\ 
 \forall \ell\in\{2,\ldots,|\cC|\}: \|P_\ell \|^2\le 4^{\beta m}\|Q_\ell \|^{-\alpha},\ K^I_\cC \le E+c\, k(m)^{-\alpha/2}\Big\}.
\end{align*}
This can be seen by comparing with the definition \eqref{H:mK} of $\cH_{m,K}$ and by noting that 
the projection \eqref{def:pro:f} of $\cH_{m,K}$ to $\widetilde{\cH}_{m,K}$ 
does not change the Jacobi coordinates.
Here $c:={n \choose 2} C_{V} C_2^{-\alpha}$ with $C_{V}$ from \eqref{eq:admiss2}.
\item 
In Corollary 6.3 of \cite{Fleischer_Knauf_2018} the following problem of integration, similar to the present
one, is considered.
It is assumed that a hypersurface $\cF\subseteq M_1\times M_2$ 
of the configuration manifold has the property that both families
\[\cF^{\,q_2}_1:=\{q_1\in M_1\mid (q_1,q_2)\in \cF\}\qquad(q_2\in M_2)\]
and
\[\cF^{\,q_1}_2:=\{q_2\in M_2\mid (q_1,q_2)\in \cF\}\qquad(q_1\in M_1)\]
consist of hypersurfaces of $M_1$ respectively of $M_2$.
Then for a classical mechanical system with Hamiltonian $H(q,p)=T_1(q,p_1)+T_2(q,p_2)+V(q_1,q_2)$
symplectic volume of a codimension two surface of phase space $T^*(M_1\times M_2)$,
is given by the sum of two integrals, involving $\cF^{\,q_2}_1$ respectively $\cF^{\,q_1}_2$.

In the present setting $M_1=\Delta^E_\cC\cap B_{R(m)}$ and $M_2=\Delta^I_\cC$.
The hypersurface $\cF$ projects to the sphere $S_{m,\cC}\subseteq M_2$, so that 
the conditions of Corollary 6.3 are violated, but in a way that the $\cF^{\,q_2}_1$ integral vanishes anyhow, 
and it is enough to treat the $\cF^{\,q_1}_2$ integral, which we will do below.
\item 
One advantage of estimating the volume of $\widehat{\cH}_{m,K}$ instead of the one of 
$\widetilde{\cH}_{m,K}$ is that the former is defined only by using absolute values of the internal
variables. Concerning the external Jacobi variables,
we denote their radii by $r_\ell:=\|Q_\ell\|$.  
Since $\widehat{\cH}_{m,K}$ is invariant with respect to rotations of the corresponding vectors,
the integration is reduced to  
\begin{align*}
\int_{\widehat{\cH}_{m,K}}\widetilde{\Omega}\ 
\le\ &
\int_{B(R(m))} dQ_1 \int_{B(2^{\beta m})} dP_1 \ 
\prod_{\ell=2}^{|\cC|} \int_{B(2R(m))} \int_{B(2^{\beta m} \|Q_l\|^{-\alpha/2}) } dP_\ell\, dQ_\ell\\ 
& \times v_{d(n-|\cC|-1)} \int_{S_{m,\cC}} ( E+c\, k(m)^{-\alpha/2} )^{(d(n-|\cC|)-1)/2} dS_{m,\cC}\\
\le\ &
c_1 2^{d\beta m} R(m)^d
\prod_{\ell=2}^{|\cC|} \Big(2^{d\beta m} \int_{\bR^d}\idty_{B(R(m))}(Q_\ell)\ v_{d}\| Q_\ell\|^{-d\alpha/2}\,dQ_\ell\Big)\\
&\times 
 \,\big( (E+c\, k(m)^{-\alpha/2})\big)^{(d (n - | \cC |) -1)/2}
\int_{S_{m,\cC}} \! dS_{m,\cC}\\
=\ & 
c_2 2^{d|\cC|\beta m} R(m)^d
\prod_{\ell=2}^{|\cC|} \Big(s_{d-1}v_d \int_{0}^{R(m)}(r_\ell)^{(d-1)(1-\alpha/2)-\alpha/2}\,dr_\ell\Big)\\
&\times
\big((E+c\, k(m)^{-\alpha/2})\big)^{(d (n - | \cC | )-1)/2}
\,k(m)^{(d (n - |\cC|)-1)/2}\\
=\ & 
c_3 2^{d|\cC|\beta m} 
R(m)^{d((1-\alpha/2)|\cC|+\alpha/2)}\\
&\times
\big(c_4(E)\big)^{(d (n - | \cC | )-1)/2}
\,k(m)^{(1-\alpha/2)(d (n - |\cC|)-1)/2}=: {\rm Int}(m),
\end{align*}
again with the volume $v_m$ of the $m$-dimensional unit ball, the surface area 
$s_m$ of the sphere $S^m$, with 
\[c_1:=2^{d(|\cC|-1)}v_d^2\,v_{d (n - |\cC|)-1}\qmbox{,}
c_2:=c_1\, s_{d (n - |\cC|)-1}\,(\delta^{|\cC|}-\delta^n)^{(d (n - |\cC|)-1)/2},\]
$c_3:=c_2\big(\frac{s_{d-1}v_d}{d(1-\alpha/2)} \big)^{|\cC|-1}$ and $c_4(E):= c + |E| k(1)^{\alpha/2}$.

We note that $k(m)$ has an exponent that decreases in $|\cC|$, whereas the exponent of $R(m)$
and the exponent linear in $\beta$ increase. Here $\cC\in \cP_0(N)$ so that $|\cC|\le n-1$.
So when we substitute $k(m)=4^{-m}$ and $R(m):=4^{mx}$ with $x\in(0,x_{\max})$
from \eqref{fixing:parameters}, we obtain a $\cC$-independent estimate when we choose 
in all three cases 
\begin{equation}
\label{d:neq:1}
|\cC|=n-1. 
\end{equation}
With $c_5(E):= c_3 (c_4(E))^{(d (n - | \cC | )-1)/2}$, we get
\begin{align}
\label{222}
{\rm Int}(m) &\le c_5(E) \  2^{ m \beta d(n-1)} 2^{mx d((1-\alpha/2)(n-1)+\alpha/2)}2^{-m(1-\alpha/2)(d-1)}.
\end{align}
\end{enumerate}
For $\beta>0$ and $x>0$ both small enough 
\[\lim_{m\to\infty} {\rm Int}(m)=0,\] 
since by assumption $\alpha<2$ and $d\ge2$.
\hfill $\Box$
\begin{Remark}[Binary and Multiple Collisions]\quad\label{rem:binary}\\
Note that $|\cC| = n-1$ in \eqref{d:neq:1} exactly corresponds to the case of a binary collision, that is, 
there is exactly one non-trivial cluster consisting of two particles. The fact that this case
corresponds to the 'worst-case'-scenario regarding the size of the integral $\,{\rm Int}(m)$ is consistent with
the heuristic consideration, based on the dimension \eqref{eq:DimDeltaE} of 
collision subspaces $\Delta_\cC^E$ in configuration space,  
that binary collisions should be the "most probable" ones.

Notice, however that the quotient of the phase space integrals of binary and of multiple collisions 
becomes independent
of the parameter $m$ as $\alpha\nearrow 2$. Then the above intuition becomes wrong.
\hfill $\Diamond$
\end{Remark}
%
\section{Time Integral of Kinetic Energy}
\label{sec:Time:Int}

Below in Proposition \ref{prop:int:kin} we will prove finiteness of the time integral of
kinetic energy. Our proof method can be considered as based on the one for Chakerian's packing theorem. 
We present here only its most basic version.
\begin{Theorem}[Chakerian's Packing Theorem \cite{Su}] \quad\\ 
For a regular curve $c\in C^2([s_0,s_1], \bR^k\setminus\{0\})$,  parameterized by arc length,
and thus of length $L(c)=s_1-s_0$ and  with curvature 
$\kappa(s) = \| \ddot{c}(s) \|$, 
\[L(c)\le  \|c(s_1)-c(s_0)\| + \int_{s_0}^{s_1}\|c(s)\|\kappa(s)\,ds .\]
\end{Theorem}
\textbf{Proof:}
$L(c)=\int_{s_0}^{s_1} \langle \dot{c}(s), \dot{c}(s)\rangle \,ds
= \langle c(s), \dot{c}(s)\rangle|_{s_0}^{s_1} 
- \int_{s_0}^{s_1} \langle c(s), \ddot{c}(s)\rangle \,ds$.
\hfill $\Box$
\begin{Remark}[Chakerian's Packing Theorem] 
One consequence is that inside a unit ball $L(c)\le2+ \int_{s_0}^{s_1}\kappa(s)\,ds$, so in order
to be long, the curve must have mean curvature larger than $1-\varepsilon$.
In spite of the simplicity of its proof, the theorem has many interesting implications, 
see {\sc Sullivan} \cite[Section 6]{Su}.
\hfill $\Diamond$
\end{Remark}

We first transfer this in Lemma \ref{lem:finite:K:int} to motion in admissible potentials 
$V\in C^2(\bR^d\!\setminus\!\{0\},\bR)$ with 
$\nabla V(q)=\cO(\|q\|^{-\alpha-1})$ ($q\to 0$) for some $\alpha \in (0,2)$ and, say
$\lim_{\|q\|\to\infty} V(q)=0$.
So we consider the Hamiltonian flow line $t\mapsto (q(t),p(t))$ with initial condition $x\in\Coll$
on phase space $P:=T^*(\bR^k\setminus\{0\})$ for 
\[H\in C^2(P,\bR)\qmbox{,} H(q,p)=\eh \|p\|^2+V(q).\]

We additionally assume that for some $C_{V}>0$
\begin{equation}
\label{grad:ineq}
\langle q,\nabla V (q)\rangle  \leq C_{V} +\alpha\, V_- (q)\qquad (\|q\| \le 1),
\end{equation}
with $V_-(q):=\max(-V(q),0)$.\footnote{This condition on $V$ corresponds to \eqref{eq:admiss2}
for $n=2$ particles, which is in that case more general than \eqref{eq:admiss1} in
Definition \ref{def:admissible}.}
Here we can simply set
\[\Coll := \big\{ x \in P \mid  \lim_{t \nearrow T(x)} q(t,x)=0\big\}.\]
\begin{Lemma}[Integral of Kinetic Energy -- Potential Scattering]\quad\label{lem:finite:K:int}\\
Under this assumption the collision trajectories have a  finite integral of kinetic energy: 
\[\int_0^{T(x)} \|p(t,x)\|^2 \,dt<\infty  \qquad (x\in \Coll).\]
\end{Lemma}
\textbf{Proof:}\\
For $\tau\in (0,T(x))$ let $\cK(\tau):=\int_0^\tau \|p(t)\|^2\,dt=\int_0^\tau \|\dot{q}(t)\|^2\,dt$. 
For $E:=H(x)$ the curve is reparameterized to unit speed by the diffeomorphism to its image
$s(t):=\int_0^t \sqrt{2(E-V(q(t')))}\,dt'$. The image equals $(0,S)$, with $S:=\lim_{t\to T(x)}s(t)\in (0,\infty]$.
So assuming without loss of generality that $\|q(t)\|\le 1$ for $t\in [0,T(x))$, with the Heaviside function
$\theta$,
\begin{align*}
\cK(\tau) &= \int_0^\tau \langle\dot{q}(t),\dot{q}(t)\rangle\,dt
= \langle q(t),\dot{q}(t)\rangle|_0^\tau- \int_0^\tau \langle q(t),\ddot{q}(t)\rangle\,dt\\
&= \langle q(t),\dot{q}(t)\rangle|_0^\tau+ \int_0^\tau \langle q(t),\nabla V(q(t))\rangle\,dt\\
&\leq \langle q(t),\dot{q}(t)\rangle|_0^\tau+ \int_0^\tau \theta\big(V(q(t)\big)\, C_{V} \,dt \\
&\qquad \qquad \qquad \
+  \int_0^\tau \theta\big(-V(q(t)\big)\,\frac{C_{V}-\alpha V(q(t))}{2(E-V(q(t)))}\langle\dot{q}(t),\dot{q}(t)\rangle \,dt\\
&\le  \langle q(t),\dot{q}(t)\rangle|_0^\tau+ (C_{V}-\alpha E)\tau+
\int_0^\tau \theta\big(-V(q(t))\big)\,\frac{\alpha}{2}\,\langle\dot{q}(t),\dot{q}(t)\rangle\,dt\\
&\le  \langle q(t),\dot{q}(t)\rangle|_0^\tau+ (C_{V}-\alpha E)\tau+
\int_0^\tau \frac{\alpha}{2}\,\langle \dot{q}(t),\dot{q}(t) \rangle\,dt.
\end{align*}
The function $\tau\mapsto \langle q(\tau),\dot{q}(\tau)\rangle$ is bounded on $[0,T(x))$,
since $\tau\mapsto \|q(\tau)\|$ is bounded and 
\[\lim_{\|q\|\searrow 0} \|q\|\,\sqrt{E-V(q)}\le \lim_{\|q\|\searrow 0}\|q\|\sqrt{E+C\|q\|^{-\alpha}}
\le \lim_{\|q\|\searrow 0}\|q\|(2C\|q\|)^{-\alpha/2} = 0.\]
So
\[\lim_{\tau\nearrow T(x)}\cK(\tau) \ \le\ 
(1-\alpha/2)^{-1} \left(\langle q(t),\dot{q}(t)\rangle|_0^{T(x)}+ (C_{V}-\alpha E)\tau\right)
\ <\ \infty.\hfill\qquad\qquad\mbox{$\Box$}\]

\begin{Remark}\quad\\[-6mm] 
\begin{enumerate}[1.]
\item 
In particular, by Cauchy's inequality, the length 
$\lim_{\tau\to T(x)} \int_0^{\tau} \|\dot{q}(t)\|\,dt$ of the collision curve is finite.
\item 
Condition \eqref{grad:ineq} is met (with $\alpha\in[\hat{\alpha},2)$ and $C_{V}:=0$) by all potentials of the form 
\[V(q)=Z\,\|q\|^{-\hat{\alpha}}\qmbox{with}\hat{\alpha}\in(0,2)\mbox{ and }Z\in\bR.\]
\item 
Condition \eqref{grad:ineq} alone, without the assumption $\nabla V(q)=\cO(\|q\|^{-\alpha-1})$,
implies that $V_-(q)=\cO(\|q\|^{-\alpha})$.\\ 
This follows by using the 
solution $W(r)=(C_V/\alpha+W(1))r^{-\alpha}-C_V/\alpha$ 
of the differential equation $r W'(r) = -C_V-\alpha W(r)$ and by integrating radially.

But \eqref{grad:ineq} is not implied by this estimate.
\item 
As the following example shows, the estimate $V_-(q)=\cO(\|q\|^{-\alpha})$ 
is not sufficient to prove the statement of
Lemma \ref{lem:finite:K:int}, even when sharpened to an estimate of the form 
$\|\nabla V(q)\|=\cO(\|q\|^{-\alpha-1})$ for some $\alpha\in (0,2)$.
\hfill $\Diamond$
\end{enumerate}
\end{Remark}
\begin{Example}[Moving on a Spiral: Optimality of Condition \eqref{grad:ineq}]\label{ex:lituus} \quad\\
The construction of the potential is based on the  {\bf  lituus spiral} 
\[c\in C^\infty([1,\infty),\bR^2)\qmbox{,}c(s)=(\cos(s),\sin(s))/s^2.\]
It converges to $\lim_{s\to+\infty}c(s)=0$, and its speed equals 
$\|c'(s)\|=\sqrt{4+s^2}/s^3$, so that it is of finite length.
The cosine of the angle between $c(s)$ and $c'(s)$ equals $-2/\sqrt{4+s^2}$; in particular
the angle converges to $\pi/2$. 

We find a more suitable (time) parameterization
$t\mapsto \tilde{c}(t):=c(s(t))$ by the assumptions, justified by 1.) and 2.) below,
\[\textstyle \|\frac{d\tilde{c}(t)}{dt}\|=\sqrt{2}s(t)\qmbox{and}s(0)=1.\]
Thus $\frac{dt}{ds}= \|c'(st)\|/ (\sqrt{2}s(t))=\sqrt{2+s(t)^2/2}\,/\,s(t)^4$, and
\[ t(s)=\frac{\left(5 \sqrt{5} s-\sqrt{s^2+4}\right) s^2-4 \sqrt{s^2+4}}{12 \sqrt{2}s^3}.\]
An asymptotic solution of the inverse near $s=\infty$ is: $s(t)\sim (2^{3/2}(T-t))^{-1/2}$, with 
collision time $T:=\left(5 \sqrt{5}-1\right)/(12\sqrt{2})$. 

We now sketch how to find a potential $V\in C^2(\bR^2\!\setminus\!\{0\},\bR)$ with the following properties.
\begin{enumerate}
\item 
Along the curve we set $V(c(s)):=-1/\|c(s)\|=-s^2$, so that for $E:=0$ the speed of a particle in $V$
with holonomic condition set by $c$ equals $\sqrt{2}s$.
\item 
In order to let the trajectory $\tilde{c}$ of the Hamiltonian $H(q,p)=\eh\|p\|^2 +V(q)$
move on the image $\hat{c}$ of $c$, the
component of $\nabla V(\tilde{c}(t))$ perpendicular to $c'(s(t))$ must be equal to
the corresponding component of the acceleration $d^2\tilde{c}/dt^2(t)$.
That component  is of size  $\frac{2 s^5 \left(s^2+2\right)}{\left(s^2+4\right)^{3/2}}$. 
\end{enumerate}
These data determine $\,V$ and its gradient along $c$, see Figure \ref{fig:spiral}.
\begin{figure}
\begin{center}
\includegraphics[width=.5\textwidth]{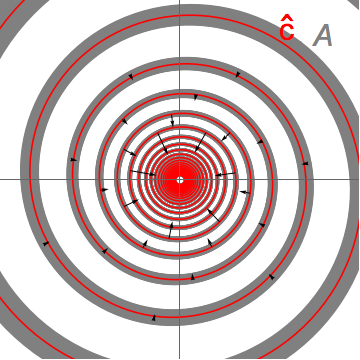}
\end{center}
\vspace*{-5mm}
\caption{Lituus spiral $\hat{c}$, its neighborhood $A$, and the force field $-\nabla V$ along the spiral}
\phantomsection \label{fig:spiral}
\end{figure}
As the component of $\nabla V$ 
perpendicular to $c'$ is non-vanishing, one can smoothly extend $V$ to a small closed 
neighborhood $A\subseteq \bR^2\!\setminus\!\{0\}$ of $\hat{c}$.
Using Tietze's extension theorem, one finds a continuous extension 
$\overline{V}\in C(\bR^2\!\setminus\!\{0\},\bR)$. Applying the theorem locally, $\overline{V}$ 
even preserves the property $\overline{V}(q)=\cO(\|q\|^{-1})$.

Then this function is smoothened on small neighborhoods of $(\bR^2\!\setminus\!\{0\})\setminus A$, 
not containing $\hat{c}$. This can be done by convolution with a smooth function, whose
support has radius $\|q\|^{-2}$, thus being smaller than the distance of neighboring segments of the
lituus spiral.

The function $V$ is then still of order $V(q)=\cO(\|q\|^{-1})$, like the Kepler potential. 

$\|\nabla V(c(s))\|$  is asymptotic to $2s^4=2\|c(s)\|^{-2}$. So by the above remark on the angle between
$c(s)$ and $c'(s)$,
\[\langle c(s),\nabla V(c(s))\rangle \sim 2 V(c(s)),\]
which means
that the assumption \eqref{grad:ineq} of Lemma \ref{lem:finite:K:int} just fails to be true.

The time integral of kinetic energy, which was finite in Lemma \ref{lem:finite:K:int} is now infinite:
\[\int_0^{T} {\textstyle \|\frac{d\tilde{c}(s(t))}{dt}\|^2 }\,dt = \int_0^{T} s^2\,dt=\int_1^\infty s \|c'(s)\|ds
= \int_1^\infty \sqrt{4+s^2}/s^2\, ds =\infty.\hfill \quad\mbox{ $\Diamond$}\]
\end{Example}
Example \ref{ex:lituus} provides the justification for condition \eqref{eq:admiss2} in our definition 
of admissible potentials.
\begin{Proposition}[Integral of Kinetic Energy -- $n$-Body Scattering]\quad\label{prop:int:kin}\\ 
For admissible potentials (see Definition \ref{def:admissible})
\begin{equation}
\label{kin:ineq}
\textstyle \int_0^{T(x)} K(p(t,x))\,dt<\infty\qquad(x\in \Coll).
\end{equation}
\end{Proposition}
\textbf{Proof:}\\
$\bullet$
The proof is similar to the one of Lemma \ref{lem:finite:K:int}. As the kinetic energy $K_{\cC_{\min}}^E$ 
of the barycenter is a constant of the motion, 
$\cK(\tau) := \int_0^\tau K_{\cC_{\min}}^I(p(t))\,dt =$ 
\begin{align}
&= 
\int_0^\tau \sum_{i\in N} \frac{m_i \| \dot{q}_i\|^2 }{2}\,dt -\frac{\|\sum_{i\in N} p_i(0)\|^2}{2m_N} \ \tau\ 
=\frac{1}{2m_N} \sum_{i<j\in N}m_i m_j  \int_0^\tau \|\dot{q}_i-\dot{q}_j\|^2 \,dt \nonumber\\
&=\sum_{i<j\in N}\frac{m_im_j}{2m_N} \langle q_i(t)-q_j(t) , \dot{q}_i(t)-\dot{q}_j(t) \rangle \big|_0^\tau
\label{first}\\
&\qquad
-\sum_{i< j\in N} \int_0^\tau \frac{m_im_j}{2m_N}\langle q_i(t)-q_j(t) , \ddot{q}_i(t)-\ddot{q}_j(t) \rangle\,dt. 
\label{second}
\end{align}
We treat the terms \eqref{first} and \eqref{second} separately. \\
$\bullet$
\eqref{first} is uniformly bounded for $\tau\in [0,T(x))$. 
To show this, we consider 
\[J_{\cC_{\min}}^I(q) = \sum_{i<j\in N}\frac{m_im_j}{m_N}\|q_i-q_j\|^2,\] 
see \eqref{Z8}. Its time derivative 
\[\frac{d}{dt}J_{\cC_{\min}}^I(q(t))
=2  \sum_{i<j\in N}\frac{m_im_j}{m_N}\big\langle q_i(t)-q_j(t) , \dot{q}_i(t)-\dot{q}_j(t)\big\rangle\]
along the orbit is four times the integrand of  \eqref{first}. 
Its second derivative equals
\[\frac{d^2}{dt^2}J_{\cC_{\min}}^I(q)
=2 K_{\cC_{\min}}^I(p)-\sum_{i<j\in N}\langle q_i-q_j , \nabla V_{i,j}(q_i-q_j)\rangle.\]
For both alternative conditions of admissibility, this is bounded below:
\begin{enumerate}[1.]
\item
Assuming \eqref{eq:admiss1},
\[\frac{d^2}{dt^2}J_{\cC_{\min}}^I(q)
\ge 2 K_{\cC_{\min}}^I(p) + \alpha \sum_{i<j\in N} \frac{Z_{i,j}} {\|q_i-q_j \|^\alpha}-{\textstyle {n\choose 2}}C_V.\]
By Remark \ref{rem:condition1} this implies
\[\frac{d^2}{dt^2}J_{\cC_{\min}}^I(q)
\ge 2 K_{\cC_{\min}}^I(p) + \alpha V(q)-C\ge\alpha E'-C,\]
with $E':=E-K_{\cC_{\min}}^E (p(0))$.
\item 
Assuming \eqref{eq:admiss2},
\begin{align*}
\frac{d^2}{dt^2}J_{\cC_{\min}}^I(q)
&\ge2 K_{\cC_{\min}}^I(p)-\alpha \sum_{i<j\in N} (V_{i,j})_-(q_i-q_j)-{\textstyle {n\choose 2}}C_V\\
&=2 K_{\cC_{\min}}^I(p)+\alpha \sum_{i<j\in N} \min\big(V_{i,j}(q_i-q_j),0\big)-{\textstyle {n\choose 2}}C_V\\
&\ge 2 K_{\cC_{\min}}^I(p)+\alpha V(	q)- {\textstyle {n\choose 2}}(C_{V}+\alpha V_{\max})\\
&\ge\alpha E' - {\textstyle {n\choose 2}}(C_{V}+\alpha V_{\max}),
\end{align*}
which is finite since by definition of admissibility
\begin{equation}
\label{V:max}
V_{\max}:=\sup\big\{V_{i,j}(q)\ \big| \ i<j\in N, \ q\in\bR^d\!\setminus\!\{0\}\big\}\in [0,\infty).
\end{equation}
\end{enumerate}
So $\liminf_{t\nearrow T(x)}\frac{d}{dt}J_{\cC_{\min}}^I(q(t))> -\infty$.\\
But as $J_{\cC_{\min}}^I(q)>0$ and $\lim_{t\nearrow T(x)} J_{\cC_{\min}}^I(q(t))=0$, we get 
\[\limsup_{t\nearrow T(x)}{\textstyle \frac{d}{dt}}J_{\cC_{\min}}^I(q(t))< +\infty,\] 
too: Between any $t\in [0,T(x))$ and $T(x)$ there exists a $t'$ with $\frac{d}{dt}J_{\cC_{\min}}^I(q(t'))\le0$.
So for $t''\in [0,T(x))$ we have 
\[\frac{d}{dt}J_{\cC_{\min}}^I(q(t'))\le-\int_{t''}^{T(x)}\frac{d^2}{dt^2}J_{\cC_{\min}}^I(q(t))\,dt
\le -(C_{V}+\alpha E')T(x).\]
$\bullet$
For \eqref{second},
\begin{align}
&-\!\!\sum_{i< j\in N} \int_0^\tau \frac{m_im_j}{2m_N}\langle q_i-q_j , \ddot{q}_i-\ddot{q}_j \rangle\,dt
\nonumber\\
&=\sum_{i< j\in N} \int_0^\tau \frac{m_i+m_j}{2m_N}\langle q_i-q_j , \nabla V_{i,j}(q_i-q_j) \rangle\,dt
\nonumber\\
&+\sum_{i< j\in N} \int_0^\tau \sum_{k\in N\setminus\{i,j\}}
\frac{\langle q_i-q_j , m_j\nabla V_{i,k}(q_i-q_k)-m_i\nabla V_{j,k}(q_j-q_k) \rangle}{2m_N}\,dt
\nonumber\\
&=\eh \sum_{i< j\in N} \int_0^\tau \langle q_i-q_j , \nabla V_{i,j}(q_i-q_j) \rangle\,dt.
\label{that:term}
\end{align}
Without loss of generality we assume $\|q_i-q_j\|\leq 1$.
\begin{enumerate}[1.]
\item 
Assuming  condition \eqref{eq:admiss1} in the definition of admissibility, \eqref{that:term} is estimated by
\begin{align*}
\eh& \sum_{i< j\in N} \int_0^\tau \langle q_i-q_j , \nabla V_{i,j}(q_i-q_j) \rangle\,dt
\le C_5\tau -\frac{\alpha}{2}\sum_{i< j\in N} \int_0^\tau \frac{Z_{i,j}}{\|q_i-q_j\|^\alpha}\,dt\\
&\le C_6\tau + \frac{\alpha}{2}\int_0^\tau [E'-V(q)] \,dt
=C_6\tau + \frac{\alpha}{2}\int_0^\tau K_{\cC_{\min}}^I (p)\,dt,
\end{align*}
with $E':=E-K_{\cC_{\min}}^E (p(0))$ and $C_6:=C_5+\frac{\alpha}{2}\big({n\choose2}C-E'\big)$,
$C$ being the constant from Remark \ref{rem:condition1}.
\item 
Similarly, for the alternative condition \eqref{eq:admiss2} of admissibility,
\begin{align*}
&\eh \!\!\! \sum_{i< j\in N} \int_0^\tau \langle q_i-q_j , \nabla V_{i,j}(q_i-q_j) \rangle\,dt
\le\eh \!\!\!  \sum_{i< j\in N} \int_0^\tau   \big[C_{V}+ \alpha\,(V_{i,j})_-(q_i-q_j)\big]\,dt\\
&= C_2\tau +\eh  \int_0^\tau \alpha\big[\sum_{i< j\in N} (V_{i,j})_-(q_i-q_j)\big]\,dt\\
&\le C_3\tau + \eh  \int_0^\tau \alpha\big[E'-V(q)\big]\,dt
\le C_3\tau + \frac{\alpha}{2} \int_0^\tau   K_{\cC_{\min}}^I (p)\,dt,
\end{align*}
with $C_{V}$ from Definition \ref{def:admissible}, $C_2:= \eh {n\choose 2} C_{V}$
and $C_3:=C_2+\frac{\alpha}{2} \big( {n\choose2}V_{\max} - E' \big)$, see \eqref{V:max}.
\end{enumerate}
$\bullet$
We arrive at an inequality of the form
\[(2-\alpha) \cK(\tau)\le C_4+C_3\tau \qquad \big(\tau\in [0,T(x))\big).\]
Since $\alpha<2$, this shows boundedness of \eqref{kin:ineq}.
\hfill $\Box$\\[2mm]
The internal cluster energy of a cluster $C\subseteq N$ equals
$H_C^I (q,p)=K_C^I(p)+V_C^I(q)$ with $K_C^I(p):=\sum_{i\in C} \frac{\|p_i^I\|^2}{2m_i}$ 
and $V_C^I(q):=\sum_{i<j\in C}V_{i,j}(q)$. Its limit value at collision time exists:
\begin{Corollary}[Collision Limits] \label{cor:coll:limit}
For $x\in \Coll$ and $\cC:={\rm SP}(x)$, 
\[\int_0^{T(x)}K_C^I(p)\,dt<\infty\qmbox{and} 
 \lim_{t\nearrow T(x)} H_C^I \big(q(t),p(t)\big)\, \in\, \bR\qquad (C\in\cC).\]
The limit $ \lim_{t\nearrow T(x)} (q_\cC^E(t),p_\cC^E(t))$ of external cluster coordinates exists.
\end{Corollary}
\textbf{Proof:}\\
$\bullet$
By Proposition \ref{prop:int:kin} the time integral of the total kinetic energy $K$ is bounded for collision
orbits. As $K=K_\cC^E+K_\cC^I$ with $K_\cC^I=\sum_{C\in \cC}K_C^I $ , 
see Lemma \ref{pro:BaryRelEnergy}, and both external and internal
kinetic cluster energies are nonnegative, the first statement follows.\\
$\bullet$
The time derivative along the orbit equals
\[\frac{d}{dt}H_C^I (q,p)= \sum_{i\in C} \left\langle \dot{q}_i^I(t),
\sum_{k\in N\setminus C}\big(-\nabla V_{i,k}(q_i-q_k)+ \sum_{j\in C}\frac{m_j}{m_C}\nabla V_{j,k}(q_j-q_k)  \big)\right\rangle.\]
The square of the first vector valued function $t\mapsto \dot{q}_i^I(t)$ is integrable, using the first statement.
The norm of the second vector valued function is uniformly bounded for $t\in [0,T(x))$.
By the Cauchy-Schwarz inequality the second statement follows.\\
$\bullet$
$\lim_{t\nearrow T(x)} q_\cC^E(t) = \lim_{t\nearrow T(x)} q(t)$ exists by definition of a collision singularity. 
The  limit of $p_\cC^E$ exists, since $\inf_{t\in [0,T(x))}\|q_i(t)-q_j(t)\| > 0$ if $[i]_\cC\neq [j]_\cC$.
So $\|\dot{p}_\cC^E(t)\|$ is bounded above on $[0,T(x))$. 
\hfill $\Box$
%
%
\section{Hitting the Poincar\'e Sections}
\label{sec:Hitting:P}
Finally we show that almost every collision orbit hits almost all Poincar\'e surfaces.

There is a natural family of Hamiltonians associated to the Jacobi map of Lemma~\ref{lem:JM}.
To simplify notation, we consider them for one arbitrary index $\ell\in \{2,\ldots,k\}$
and henceforth omit that index.
Then (as in Remark \ref{rem:J:differences}) we have set partitions $\cC\preccurlyeq\cD$  
with $\ell=|\cC|=|\cD|+1$, 
and there is a unique cluster $D_U\in\cD$ which is the disjoint
union $C_L\ \dot{\cup}\ C_R$ of two clusters $C_L,C_R\in \cC$, 
and the other clusters of $\cD$ coincide with the other clusters of $\cC$.
The difference of the external Hamiltonians from Lemma \ref{pro:BaryRelEnergy} is
\[ {\frak H} := H^E_\cC-H^E_\cD\ =\ K^E_\cC-K^E_\cD \ + \ V^E_\cC-V^E_\cD.\]
Here, with reduced mass
$\mu:= \frac{ m_{C_L} m_{C_R}}{ m_{D_U}}$ and Jacobi coordinates 
\[(Q,P)=
\left(q_{C_L}-q_{C_R},\ 
\mu\left({\textstyle \frac{P_{C_L}}{m_{C_L}}-\frac{P_{C_R}}{m_{C_R}} }\right)\right)\in T^*\bR^d\]
\begin{align}
K^E_\cC(p)-K^E_\cD(p) &=\sum_{C\in \cC}\frac{\|p_C\|^2}{2m_C} - \sum_{D\in \cD}\frac{\|p_D\|^2}{2m_D} 
= \frac{\|p_{C_L}\|^2}{2m_{C_L}}+\frac{\|p_{C_R}\|^2}{2m_{C_R}} - \frac{\|p_{D_U}\|^2}{2m_{D_U}}
\nonumber \\ 
\label{P:l}
&= \frac{\|P\|^2}{2\mu}.
\end{align}
\begin{align*}
V^E_\cC(q)-V^E_\cD(q) &=  
\sum_{i<j\in N:\, [i]_\cC \neq [j]_\cC} V_{i,j}(q_i - q_j)-\sum_{i<j\in N:\, [i]_\cD \neq [j]_\cD} V_{i,j}(q_i - q_j)\\
&=\sum_{i\in C_L,\, j\in C_R}V_{i,j}(q_i - q_j)
=\sum_{i\in C_L,\, j\in C_R}V_{i,j}\big(Q+\big(q^I_\cC\big)_i  - \big(q^I_\cC\big)_j\big).
\end{align*}

We now consider the change of the Jacobi kinetic energy \eqref{P:l} along a solution curve
by the fundamental theorem of calculus:
\begin{equation}
\label{f:t:c}
\frac{\|P(t)\|^2}{2\mu}= \frac{\|P(0)\|^2}{2\mu}+\int_0^t\frac{d}{d\tau} \frac{\|P(\tau)\|^2}{2\mu}\,d\tau.
\end{equation}
Using \eqref{P:l}:
\begin{align}
\frac{d}{dt}\frac{\|P\|^2}{2\mu}
&=\frac{d}{dt}\left(
\frac{\|p_{C_L}\|^2}{2m_{C_L}}+\frac{\|p_{C_R}\|^2}{2m_{C_R}} - \frac{\|p_{D_U}\|^2}{2m_{D_U}}\right)
\nonumber\\
&= \langle \dot{q}_{C_L},\dot{p}_{C_L} \rangle + \langle \dot{q}_{C_R},\dot{p}_{C_R} \rangle-
\langle \dot{q}_{D_U},\dot{p}_{D_U} \rangle \nonumber\\
&=
-\hspace*{-3mm}\sum_{i\in C_L,\, j\in N\setminus C_L}
\left\langle \dot{q}_{C_L},\,\nabla V_{i,j}(q_i - q_j)\right\rangle
-\hspace*{-3mm}\sum_{i\in C_R,\, j\in N\setminus C_R}
\left\langle \dot{q}_{C_R},\,\nabla V_{i,j}(q_i - q_j)\right\rangle \nonumber\\
&\quad +\hspace*{-3mm}\sum_{i\in D_U,\, j\in N\setminus D_U}
\left\langle \dot{q}_{D_U},\,\nabla V_{i,j}(q_i - q_j)\right\rangle.
\label{terms:123}
\end{align}
The three terms have a similar form. We show explicitly how to estimate the first term:
\begin{align}
-\hspace*{-3mm}&\sum_{i\in C_L,\, j\in N\setminus C_L}
\left\langle \dot{q}_{C_L},\,\nabla V_{i,j}(q_i - q_j)\right\rangle
=- \hspace*{-3mm} \sum_{C\in \cC\setminus \{C_L\}}\sum_{i\in C_L,\,j\in C}
\left\langle \dot{q}_{C_L},\,\nabla V_{i,j}(q_i - q_j)\right\rangle\nonumber\\
&=- \frac{d}{dt}W_L(q) +
\hspace*{-3mm} \sum_{C\in \cC\setminus \{C_L\}}\sum_{i\in C_L,\,j\in C}
\left\langle \dot{q}_{C_L},\,\nabla V_{i,j}(q^E_i - q^E_j)-\nabla V_{i,j}(q_i - q_j)\right\rangle,
\label{d:dt:Wl}
\end{align}
with 
\begin{equation}
\label{W:L}
W_L(q(t)):= \sum_{C\in \cC\setminus \{C_L\}}\sum_{i\in C_L,\,j\in C}V_{i,j}(q^E_i (t)- q^E_j(t)).
\end{equation}
By \eqref{D2V} the contribution of the first term in \eqref{d:dt:Wl} to \eqref{f:t:c} is of order
\begin{equation}
\label{WLWL}
W_L(q(0))-W_L(q(t))=\cO\left(\sum_{C\in \cC\setminus \{C_L\}}\|q_{C_L}(t)-q_C(t)\|^{-\alpha}\right)
\quad\big(t\in [0,T(x))\big).
\end{equation}
Up to now we did not pose an assumption about the location of the trajectory in configuration space.
This, however will be needed in the proof of the following proposition.

\begin{Proposition}[Hitting the Poincar\'e Surfaces]\quad\label{prop:hit}\\
For every energy $E\in\bR$ and $\sigma_E$-almost every initial condition 
$x\in \Coll$ there is an $M(x)\in \bN$ so that, with $\cH_m^\pm$ defined in \eqref{H:m:K:pm}, 
\[\Phi\big([0,T(x)),x\big)\cap \cH_m^-\neq \emptyset\qquad (m\ge M(x)).\]
\end{Proposition}
\textbf{Proof:}\\
According to Lemma \ref{lem:aa:Fm}, the forward configuration space trajectory $q([0,T(x)),x)$ intersects 
all but finitely many hypersurfaces $\cF_m$.
By Lemma 3.1 of \cite{Fleischer_Knauf_2018} we can assume that the intersections of the
forward configuration space trajectory $q([0,T(x)),x)$ with the hypersurfaces $\cF_m$ is transversal.
By transversality and closedness of $ \cup_{m\in\bN} \cF_{m} \subseteq \hM$, the set 
\[\{t\in [0,T(x))\mid q(t,x)\in \cup_{m\in\bN} \cF_{m}\}\]
is discrete in $[0,T(x))$. When we enumerate it in ascending order by $(t_i)_{i\in\bN}$, then 
$\lim_{i\to\infty }t_i=T(x)$. As the $\cF_m$ are mutually disjoint, 
the indices $m_i\in\bN$ are uniquely defined by $q(t_i,x)\in \cF_{m_i}$.

As $\cP_0(N)$ is finite,  there
is at least one $\cD\in \cP_0(N)$ with $q(t_i,x)\in \cF_{m_{i},\cD}$ for infinitely many $i\in\bN$.
As stated in Lemma \ref{lem:finer:coarser}, $\cD\preccurlyeq{\rm SP}(x)$ for those $\cD$.

In fact, it suffices to show that for any $\cD\in \cP_0(N)$ there is a $I(\cD)\in \bN$ such that 
\begin{equation}
\label{hit:HmD}
\Phi(t_i,x)\in \cH^-_{m_i,\cD}\qquad \big(i\ge I(\cD):q(t_i,x)\in \cF_{m_i,\cD}\big).
\end{equation}
Then with $I:=\max_{\cD\in \cP_0(N)}I(\cD)$ and $M(x):= m_I$ the proposition is true.

Obviously we have to prove \eqref{hit:HmD} only for those $\cD\in \cP_0(N)$ with an infinite number
of collisions with hypersurfaces $\cF_{m_i,\cD}$.

The total momentum $P_1$ (see Lemma \ref{lem:JM}) is conserved, so that the first condition in the definition
\eqref{H:mK} of the Poincar\'e surfaces $\cH_{m,K}$ is met for $m$ large.

We first consider the case $\cD = {\rm SP}(x)$.
Then not only the external positions $q_D^E$ but also the momenta $p_D^E$
for the clusters $D\in \cD$ have collision limits, see
Corollary \ref{cor:coll:limit}. So also the Jacobi coordinates $Q_\ell$ and $P_\ell$ ($\ell\in\{2,\ldots,k\}$)
from Lemma \ref{lem:JM} have limits, and the second condition in \eqref{H:mK} is met, too for $m$ large.

The second case to consider is $\cD\precneqq{\rm SP}(x)$. The norms of the Jacobi momenta $P_\ell$
are estimated based on \eqref{f:t:c}:
\[\|P_\ell(t)\|^2=\|P_\ell(0)\|^2+2\int_0^t \langle P_\ell(\tau), \textstyle \frac{d}{d\tau}P_\ell(\tau)\rangle d\tau.\]
We now consider a segment of the trajectory, with $q_\cC^E(t)$ in the Jacobi space $\cJ_K$.
Then \eqref{WLWL} can be bounded by
\[W_L(q(0))-W_L(q(t))=\cO\left(\|Q_k\|^{-\alpha}\right) ,\]
by \eqref{V:ineq} and the definition \eqref{Jacobi:space} of $\cJ_K$.

To estimate the second term
\begin{equation}
\label{second:term}
\sum_{C\in \cC\setminus \{C_L\}}\sum_{i\in C_L,\,j\in C}
\left\langle \dot{q}_{C_L},\,\nabla V_{i,j}(q^E_i - q^E_j)-\nabla V_{i,j}(q_i - q_j)\right\rangle
\end{equation}
in \eqref{d:dt:Wl}, we note that, by assumption \eqref{D2V} on the potential
\[ \|\nabla V_{i,j}(q^E_i - q^E_j)-\nabla V_{i,j}(q_i - q_j)\| \le C \|q^E_i - q^E_j\|^{-\alpha-2}\|\, (\|q^I_i\| + \|q^I_j\|).\]
But as we assumed that 
$[i]\neq [j]$,
by Lemma \ref{lem:intra:inter} the internal cluster coordinates are bounded by
\[\|q^I_i\| + \|q^I_j\| = \cO(\|q^E_i - q^E_j\|).\]
So
\begin{equation}
\label{nabla:VV}
\|\nabla V_{i,j}(q^E_i - q^E_j)-\nabla V_{i,j}(q_i - q_j)\| \le C \|q^E_i - q^E_j\|^{-\alpha-1}.
\end{equation}
Proposition \ref{prop:int:kin} states that the time integral of kinetic energy is finite. So in particular,
using the Cauchy-Schwarz inequality, the length 
$\int_0^{T(x)} \|\dot{q}_{C_L}(t)\|\,dt$ of the collision curve is finite, i.e., we can reparameterize it
by arc length, with a bounded parameter interval. By integrating \eqref{nabla:VV}, we obtain
the estimate $\cO\left(\|Q_{\ell}(t)\|^{-\alpha}\right)$ for \eqref{second:term}, too.

The other two terms in \eqref{terms:123} are estimated by the same method.
\hfill $\Box$\\[2mm]
{\bf Proof of Theorem \ref{thm:ImprobColl}:}\\
We adapt Theorem A of \cite{Fleischer_Knauf_2018} to the notation of the present article. 
Then it states under the assumptions
\begin{enumerate}
\item 
that the vector field $X_H$, restricted to the energy surface $\Sigma_E$, 
is transverse to  the Poincar\'e surfaces $\cH_m^-$,
\item 
and that these have finite volume, with $\lim_{m\to\infty}\int_{\cH_m} \Omega =0$,
\end{enumerate}
that the Liouville measure
\[\sigma_E(\Trans_E\cap \Wand_E)=0\, ,\]
with
\begin{enumerate}
\item [3.]
$\Trans_E:=
\{x\in \Sigma_E \mid \exists\, m_0\in\bN\;\forall \,m\ge m_0: \cO^+(x)\cap \overline{\cH}_m\neq\emptyset \}$, 
and
\item [4.]
the wandering set $\Wand_E$
consisting of those $x\in \Sigma_E$ which have a neighborhood   $U_x$ 
so that for a suitable time $t_-\in(0,T(x))$
\begin{equation*}
U_x\cap \Phi\big(\left((t_-,T(x))\times U_x\right)\cap D\big)=\emptyset.
\end{equation*}
\end{enumerate}
This implies that $\sigma_E(\Coll_E)=0$, since:
\begin{enumerate}
\item 
Concerning Condition 1.\ above,  Lemma \ref{lem:pa:submani} states that 
the Hamiltonian vector field $X_H$ is transverse to all $\cH_m^-$. 

\item 
The $2(nd-1)$--form $\Omega=\Omega_{nd-1}$, 
defined in \eqref{Omega} is invariant under the flow generated by $X_H$. 
As proven in (6.1) of \cite{Fleischer_Knauf_2018}, $i_{X_H}\sigma_E$ equals (possibly up to sign)
the pullback $\imath_E^*\Omega$ of that form to the energy surface $\Sigma_E$.

So Condition 2.\ has been shown in Lemma \ref{lem:pa:submani} and Proposition \ref{prop:area}.
\item 
Then Proposition \ref{prop:hit} states that $\Coll_E\subseteq \Trans$. 
\item 
As proven in Lemma 1.2 of
\cite{Fleischer_Knauf_2018}, $\Sing_E\subseteq \Wand_E$. So by Definition \eqref{eq:Coll}, 
$\Coll_E\subseteq \Wand_E$, too.
\hfill $\Box$
\end{enumerate}

\section{Addenda}
\label{sec:Addenda}
\paragraph{Collisions on the Line}\quad\\
Up to this point, we have used the assumption $d\geq 2$ only in the proof of Proposition \ref{prop:area}. 
So the same reasoning can be applied when considering collision orbits in $d=1$ dimension. 
Obviously, binary collisions cannot be improbable in this case. 
This is consistent with the last exponent in \eqref{222} becoming zero for $d=1$ and $|\cC|=n-1$ 
(which exactly corresponds to a single binary collision). 
However, if we choose $|\cC|\le n-2$ instead of \eqref{d:neq:1}, that exponent is negative
even for $d=1$. This is the case for every multiple collision configuration, that does not consist of merely one binary. Thus we have also shown:

\begin{Theorem}\quad\\
In $d=1$ dimension, for any energy $E\in \bR$
the set of all non-binary collision orbits (i.e.\ collisions consisting of a triple or higher order collision, of several collisions occurring simultaneously, or a combination thereof) has Liouville measure zero.
\end{Theorem}

\paragraph{Collisions in Systems with Centers}\quad\\
In Theorem \ref{thm:ImprobColl} the potential $V$ was assumed to be of the form \eqref{eq:DefV}.
Only minor changes need to be implemented in the presence of fixed centers, that is for potentials of
the form
\[V(q)+\sum_{i\in N,\, k=1,\ldots, k_{\max}} W_{i,k}(q_i-c_k), \]
with $W_{n,k}\in C^2(\bR^d\!\setminus\!\{c_k\},\bR)$ admissible in the sense of Definition \ref{def:admissible}.

When defining the coordinates, we have to discriminate whether one of the centers is part of the cluster or not.
In the first case, internal cluster coordinates are replaced by the particles' distances to the center, whereas the respective external coordinates are dropped (the estimates on the external coordinates' insignificant contribution to phase space volume vanish into the obvious fact that the centers don't contribute to that at all).
Otherwise, the coordinates are defined as before.
Then the Poincar\'{e} surfaces are also defined as before.
The rest is a straightforward calculation along the lines of the previous sections.

\end{document}